\documentclass[12pt]{amsart}
\usepackage{amsbsy,amssymb,amsmath,amsthm,amscd,amsfonts,latexsym,amstext,delarray,
amsmath,graphicx} 
\usepackage[margin=1in]{geometry}
\usepackage{color}
\input xypic

\usepackage[all]{xy}
\usepackage{setspace}

\newtheorem{thm}{Theorem}[section]
\newtheorem{prop}[thm]{Proposition}
\newtheorem{cor}[thm]{Corollary}
\newtheorem{lem}[thm]{Lemma}

\newtheorem{defn}[thm]{Definition}
\newtheorem{rem}[thm]{Remark}

\numberwithin{equation}{section}

\def\Z{{\mathbb Z}}
\def\N{{\mathbb N}}
\def\R{{\mathbb R}}
\def\C{{\mathbb C}}

\def\GL{{\rm GL}}

\def\cA{{\mathcal A}}
\def\cB{{\mathcal B}}

\def\cD{{\mathcal D}}
\def\cE{{\mathcal E}}

\def\cG{{\mathcal G}}
\def\cH{{\mathcal H}}
\def\cI{{\mathcal I}}

\def\cL{{\mathcal L}}
\def\cM{{\mathcal M}}
\def\cN{{\mathcal N}}

\def\cP{{\mathcal P}}

\def\cR{{\mathcal R}}
\def\cS{{\mathcal S}}
\def\cT{{\mathcal T}}
\def\cU{{\mathcal U}}
\def\cV{{\mathcal V}}
\def\cW{{\mathcal W}}

\def\bS{{\mathbb S}}

\def\bU{{\mathbb U}}

\def\Tr{{\rm Tr}}
\def\Spec{{\rm Spec}}
\def\m{{\mathfrak m}}

\def\fh{{\mathfrak h}}
\def\ff{{\mathfrak f}}
\def\fg{{\mathfrak g}}

\DeclareMathOperator*{\Res}{Res}

\def\cancel#1#2{\ooalign{$\hfil#1\mkern1mu/\hfil$\crcr$#1#2$}}
\def\Dirac{\mathpalette\cancel D}

\def\cutint{{\int \!\!\!\!\!\! -}}

\title{Spectral Action Models of Gravity on Packed Swiss Cheese Cosmology}
\author{Adam Ball and Matilde Marcolli}
\address{Division of Physics, Mathematics and Astronomy, 
Caltech, 1200 E. California Blvd. Pasadena, CA 91125, USA}
\email{aaball@caltech.edu}
\email{matilde@caltech.edu}

\date{}

\begin{document}
\maketitle

\begin{abstract}
We present a model of (modified) gravity on spacetimes with fractal structure
based on packing of spheres, which are (Euclidean) variants of the Packed 
Swiss Cheese Cosmology models. As the action functional for gravity we
consider the spectral action of noncommutative geometry, and we compute
its expansion on a space obtained as an Apollonian packing of
$3$-dimensional spheres inside a $4$-dimensional ball. Using information
from the zeta function of the Dirac operator of the spectral triple, we compute
the leading terms in the asymptotic expansion of the spectral action. They
consist of a zeta regularization of a divergent sum which involves the leading terms
of the spectral actions of the individual spheres in the packing. This 
accounts for the contribution of the points $1$ and $3$ in the dimension
spectrum (as in the case of a $3$-sphere). There is an additional term 
coming from the residue at the additional point in the real dimension spectrum 
that corresponds to the packing constant, as well as a series of 
fluctuations coming from log-periodic oscillations, created by the points
of the dimension spectrum that are off the real line. These terms detect the 
fractality of the residue set of the sphere packing. We show that the presence of
fractality influences the shape of the slow-roll potential for inflation, 
obtained from the spectral action. We also discuss the effect of truncating
the fractal structure at a certain scale related to the energy scale in the
spectral action. 
\end{abstract}

\section{Introduction}

\subsection{Fractal structures in cosmology}

The usual assumptions of isotropy and homogeneity of spacetime would
require that the matter distribution scales uniformly in space.  Large
scale violations of homogeneity were discussed, for instance, in \cite{Rees},
while the idea of a fractal distribution of matter, scaling with a fractal dimension
$D  \neq 3$, was suggested in \cite{Peeb}. More recently,
a growing literature based on the analysis of redshift catalogs at the level of
galaxies, clusters, and superclusters has collected considerable evidence for
the presence of fractality and multifractality in cosmology. We refer the
reader to the survey \cite{Sylos} for a detailed discussion, see also \cite{Sylos2}.
While there is no complete agreement on the resulting dimensionality, partly
due to difficulties in the interpretation of redshift data in estimating co-moving 
distances, multifractal models in cosmology have been widely studied in 
recent years. Cosmological models exhibiting a fractal structure  
can be constructed, adapting the original ``swiss cheese model" of \cite{Rees}.
The resulting models are usually referred to as Packed Swiss Cheese Cosmology 
(PSCC), see \cite{MuDy} for a recent detailed survey. 
The main idea in the construction of swiss cheese models of cosmology 
is to have spacetimes that are locally inhomogenous but appear globally 
isotropic and that satisfy everywhere the Einstein equation. In the original
construction of PSCC models, in a region
defined by a standard Friedmann--Robertson--Walker (FRW) cosmology, several
non-overlapping spheres are inscribed, inside which the mass is contracted
to a smaller higher density region, hence creating inhomogeneities. The solution
inside the ball is patched to the external FRW solution along a surface with
vanishing Weyl curvature tensor (which ensures isotropy is preserved). A
swiss cheese type model based on the Tolman metric was developed in \cite{Rib1},
\cite{Rib2}. In Packed Swiss Cheese Cosmology models, a configuration of such spheres is
chosen so that they are tangent to each other and arranged into a higher dimensional
version of the Apollonian packing of circles. In a variant of this model, see the
discussion in \S 8 of \cite{MuDy}, instead of compressing the matter inside each
spherical region, at each stage of the construction process the matter is expanded
to lie along the spherical shell, so that one ends up with a model of gravity
interacting with matter, supported on the resulting fractal. The point of view we
follow in this paper is similar to the latter: we consider spacetimes that
are products of a time direction and a fractal arrangement of $3$-spheres
(or of other spherical space forms). We develop a model of gravity on such 
Packed Swiss Cheese Cosmology (PSCC) models using the spectral action
as an action functional for gravity.

\subsection{Spectral triples}
In Noncommutative Geometry, the formalism of {\em spectral triples} extends
ordinary Riemannian (and spin) geometry to noncommutative spaces, \cite{CoS3}. This approach encodes
the metric structure in the data of a triple $ST=(\cA,\cH,D)$ of an involutive algebra $\cA$
(associative, but not necessarily commutative), with a (faithful) representation $\pi: \cA \to \cB(\cH)$ 
by bounded operators on a Hilbert space $\cH$, and with the additional structure of a {\em
Dirac operator} $D$, namely an unbounded, self-adjoint operator, densely defined on $\cH$
with the properties:
\begin{description}
	\item[(i)] $(I+D^2)^{-1/2}$ is a compact operator
	\item[(ii)] for all $a\in \cA$, the commutators $[D,\pi(a)]$ are densely 
		defined and extend to bounded operators on $\cH$.
\end{description}
The metric dimension of a spectral triple is defined as
\begin{equation}\label{metdim}
 \mathfrak{d}_{ST} := \inf \{ p > 0 \mid \text{tr}((I+D^2)^{-p/2} < \infty \} . 
\end{equation} 
A spectral triple is said to be finitely summable if $\mathfrak{d}_{ST}<\infty$.

\smallskip

The notion of dimension for a spectral triple is more elaborate than just the
metric dimension. Indeed, a more refined notion of dimension is given by the
{\em dimension spectrum}, $\Sigma_{ST} \subset \C$. This is a set of complex
numbers, defined as the set of poles of a family of zeta functions associated to
the Dirac operator of the spectral triple. In the case where ${\rm Ker}\, D=0$,
the zeta function of the Dirac operator is given by $\zeta_D(s)=\Tr(|D|^{-s})$.
Let $\delta(T)=[|D|,T]$ and let $\cB$ denote the algebra generated by the $\delta^m(\pi(a))$
and $\delta^m([D,\pi(a)])$, for all $a\in \cA$, and $m\in \N$. One considers additional
zeta functions of the form $\zeta_{D,a}(s)=\Tr(a|D|^{-s})$, for arbitrary $a\in \cA$
and $\zeta_{D,b}(s)=\Tr(b|D|^{-s})$, for arbitrary $b\in \cB$. The dimension
spectrum is the set of poles of the functions $\zeta_{D,a}(s)$ and $\zeta_{D,b}(s)$.
It typically includes other points, in addition to the metric dimension, and may
include real non-integer points as well as complex points off the real line.
Spectral triples associated to fractals typically have non-integer and non-real
points in their dimension spectrum. In the following, we will use the notation
$\Sigma_{ST}^+ :=\Sigma_{ST}\cap \R_+$ for the part of the dimension
spectrum contained in the non-negative real line. Geometrically, the
dimension spectrum represents the set of dimensions in which the space 
manifests itself, when viewed as a noncommutative 
space. Even in the case of an ordinary manifold, the dimension spectrum
contains additional points, besides the usual topological dimension.
The non-negative dimension spectrum $\Sigma_{ST}^+$, in particular, describes 
the dimensions that contribute terms to the action functional for gravity, as
we discuss more in detail in \S \ref{SacSec} below, %
while the points of the
dimension spectrum that lie off the real line contribute fluctuations in the
form of log oscillatory terms, as we will see in \S \ref{OscSec}. We say that
the dimension spectrum is {\em simple} if the poles are simple poles. Spectral
triples with simple dimension spectrum are sometimes referred to as ``regular". However,
the terminology ``regular spectral triple" is often used in the literature with a 
different meaning, related to ``smoothness" properties (see for instance \cite{Rennie}).
Thus, in the following we will use the terminology ``simple dimension
spectrum" to avoid confusion.

\smallskip

A compact spin Riemannian manifold $M$ can be described by a spectral triple
$ST_M=(C^\infty(M), L^2(M,\bS), \Dirac_M)$, 
by taking $\cA=C^\infty(M)$, the algebra of smooth functions, $\cH=L^2(M,\bS)$
the Hilbert space of square-integrable spinors, and $D=\Dirac_M$ the 
Dirac operator, which is a self-adjoint square root of the (negative) Laplacian 
of the manifold. The metric dimension of $ST_M$ agrees with the dimension of $M$,
by Weyl's law for the Dirac spectrum. One can also recover the geodesic 
distance on $M$ from $ST_M$: for any two points $x,y \in M$ 
$$d_{geo}(x,y) = \sup \{|f(x)-f(y)| \mid ||[D,\pi(f)]|| \le 1 \}.$$
A reconstruction theorem \cite{Co-rec} moreover shows that
the manifold $M$ itself can be reconstructed from the data of 
a commutative spectral triple that satisfies a list of additional axioms
describing properties of the geometry such as orientability, Poincar\'e
duality, etc. The non-negative dimension spectrum $\Sigma_{ST_M}^+$
consists of non-negative integers less than or equal to $\dim(M)$ (see \S \ref{DimSpSec}
for more details).

\subsection{The spectral action as a model for (modified) gravity}\label{SacSec}

The formalism of spectral triples plays a crucial role in the construction
of models of gravity coupled to matter based on noncommutative geometry. 
The main ideas underlying the construction of these models can be
summarized as follows: 
\begin{itemize}
\item The spectral action is a natural action functional for gravity on any (commutative or noncommutative) space described by a finitely summable spectral triple.
\item On an ordinary manifold, the asymptotic expansion of the spectral action
recovers the usual Einstein--Hilbert action of gravity, with additional modified gravity
terms (Weyl conformal gravity, Gauss--Bonnet gravity).
\item In the case of an ``almost commutative geometry" (locally a product $M\times F$
of an ordinary manifold $M$ and a finite noncommutative space) the model of gravity
on $M\times F$ given by the spectral action describes gravity coupled to matter on $M$,
with the matter content (fermions and bosons) completely determined by the geometry of 
the finite noncommutative space $F$.
\end{itemize}
We refer the reader to the detailed account of the construction of such
models given in \cite{CCM} and in Chapter 1 of \cite{CoMa-book}.
For a finitely summable spectral triple, the spectral action functional \cite{CC}
is defined as 
$$\cS(\Lambda) = \text{Tr}(f(D/\Lambda)) = 
\sum_{\lambda \in \text{Spec}(D)}\text{Mult}(\lambda)f(\lambda/\Lambda), $$
where $f$ is a non-negative even smooth approximation to a 
cutoff function and $\Lambda$ is a positive real number. 
As $\Lambda$ grows, more rescaled eigenvalues of the form $\lambda/\Lambda$ 
escape the cutoff of $f$ and the expression grows. 

\smallskip

In the case of a finitely
summable spectral triple with dimension spectrum consisting of simple
poles on the positive real line, the spectral action can be expanded asymptotically for large $\Lambda$, 
\cite{CC}. The asymptotic expansion relies on the Mellin transform
relation between the zeta function of the Dirac operator and the heat kernel.
The asymptotic expansion of the spectral action is then of the form 
\begin{equation}\label{SAexpand}
 \Tr (f(D/\Lambda))\sim\,\sum_{\beta \in \Sigma^+_{ST}}\,f_\beta\,\Lambda^\beta \,\,  
 \cutint |D|^{-\beta} \,\, +\,f(0)\,\zeta_{D}(0),
\end{equation}
where $f_\beta=\,\int_{0}^{\infty}f(v)\,v^{\beta-1}\,dv$ are the momenta of $f$, 
the summation is over the points of the non-negative dimension spectrum 
$\Sigma^+_{ST}$, and the coefficients are residues of the zeta function,
\begin{equation}\label{cutint}
 \cutint |D|^{-\beta} =  \frac{1}{2} {\rm Res}_{s=\beta}  \,\, \zeta_D(s), 
\end{equation} 
representing the noncommutative integration in dimension $\beta$.

\smallskip

In the case of a $4$-dimensional manifold $M$, one can write 
the asymptotic expansion in the form \cite{CC-uncanny}
$$\text{Tr}(f(D/\Lambda)) \sim 2\Lambda^4f_4a_0 + 2\Lambda^2f_2a_2 + f_0 a_4 ,$$
where the $f_i$ are momenta of the cutoff function $f$, with $f_0=f(0)$ and
$f_k = \int_{0}^{\infty}f(v)\,v^{k-1}\,dv$. Physically, the coefficients 
$a_0$, $a_2$ and $a_4$ correspond, respectively, to the cosmological term,
the Einstein--Hilbert term, and the modified gravity terms (Weyl curvature and 
Gauss--Bonnet) of the gravity action functional. 
In the case of an almost-commutative geometry,
the asymptotic expansion of the spectral action delivers additional {\em bosonic} terms, including
Yang--Mills terms for the gauge bosons, and kinetic and interaction terms for Higgs bosons,
and (non-minimal) coupling of matter to gravity (with the Higgs conformally coupled
to gravity). The fermionic terms in the action functional for gravity coupled to matter
come from an additional term not included in the spectral action, which accounts for 
the kinetic terms of the fermions and the boson-fermion interaction terms, see \cite{CCM}, \cite{CoMa-book}.
For the purpose of the present paper, we are only interested in the gravitational terms,
though couplings to matter could also be included, by taking a product of the
geometries we will be discussing with a finite noncommutative geometry. %

\smallskip

We will see in the next section that, in the case of the Packed Swiss Cheese
Cosmology, the spectral action has new contributions that arise from an
additional point in the dimension spectrum that reflects the fractality of the
model, as well as log-periodic oscillations contributed by the points
of the dimension spectrum that are off the real line. 

\smallskip
\subsection{Summary of results}

The main new results in this papers are structured as follows. 

\smallskip

In \S \ref{ApolloSec} we obtain an estimate, in the form of an upper bound, on 
the exponent of convergence of the zeta function $\zeta_{\cL}(s)$ of the length 
spectrum of an Apollonian packing $\cP$ of $3$-sphere (Proposition \ref{approxsigma}),
we describe the spectral triple of $\cP$ (Definition \ref{S3defPD}),
and we compute the zeta function $\zeta_{\cD_\cP}(s)$ of the Dirac operator of the 
spectral triple (Proposition \ref{zetaST}),
in terms of the zeta function of the unit $3$-sphere and the zeta function $\zeta_{\cL}(s)$
of the length spectrum. We discuss the structure of the dimension spectrum (Lemma \ref{DimSpPSC}).

\smallskip

In \S \ref{SAPSCsec}, we use the results on the zeta function to obtain an expansion
of the spectral action functional. In \S \ref{OscSec} we discuss how the heat kernel
expansion, and consequently the expansion of the spectral action, is altered by the
presence of complex points of the dimension spectrum off the real line. For the case
of a fractal geometry with exact self-similarity realized by a single contraction ratio,
we obtain an explicit form of the log-oscillatory terms coming from the non-real
points of the dimension spectrum, in the form of a Fourier series that converges
to a smooth function (Proposition \ref{singlescale}). In \S \ref{ApproxSec} we
discuss approximations by truncation of the Fourier series of the oscillatory terms.
We then identify a set of four analytic conditions on the zeta function $\zeta_\cL(s)$
of the length spectrum of the Apollonian packing (Definition \ref{defan}), which 
ensure that the spectral action
has an expansion where the oscillatory terms can be approximated by a series
of contributions from length spectra (fractal strings) with exact self-similarity. The contribution
from the real points of the dimension spectrum yields gravitational terms as in the
case of a $3$-dimensional geometry, with an additional term coming from the
only real pole of $\zeta_\cL(s)$ at its exponent of convergence (Proposition \ref{SpActPSC}).
We also compute the form of the expansion of the spectral action when taking a
geometry that is a product of the Apollonian arrangement $\cP$ of $3$-spheres
with a compactified time axis (Proposition \ref{SAprodS1P}).

\smallskip

In \S \ref{FractalScaleSec} we investigate the effect on the spectral action functional
of a truncation of the fractal structure at a certain energy dependent scale. We
obtain estimates on the size of the error term and its dependence on
the energy $\Lambda$ (Propositions \ref{alphaLambda2} and \ref{errorM}).

\smallskip

In \S \ref{DodeSec} we construct another model of fractal space, which
allows for the presence of ``cosmic topology". This is obtained by
taking a Sierpi\'{n}ski fractal arrangement of spherical dodecahedra and
then simultaneously closing up all of them via the action of the icosahedral
group, obtaining a fractal arrangement of Poincar\'e homology spheres
(usually referred to as dodecahedral spaces in the cosmic topology literature). 
This is a simpler fractal than the Apollonian sphere packing, since it has exact
self-similarity with a single contraction ratio $(2+\phi)^{-1}$, where $\phi$ is the
golden ratio. In this case we can compute more explicitly the 
new terms that arise in the expansion of the spectral action, including
the oscillatory terms (Propositions \ref{S3dodeca} and \ref{SAdodeca} and
Corollary \ref{PY4dim}).

\smallskip

In \S \ref{InflSec} we compute the effect of the additional terms in the
spectral action expansion on the shape of the slow-roll potential obtained
by perturbing the Dirac operator by a scalar field (Propositions \ref{prodS1PV}
and \ref{prodS1PVflu}).

\medskip
\section{Spectral triples and zeta functions for Packed Swiss Cheese Cosmology}\label{S3PSCsec}

\medskip
\subsection{Apollonian packings of $D$-dimensional spheres}\label{ApolloSec}

Higher dimensional generalizations of the Apollonian packings of circles
in the plane, consisting of ``packings" of $(D-1)$-dimensional hyperspheres
$S^{D-1}$ inside a $D$-dimensional space $\R^D$, were variously studied, for
instance in \cite{Farr}, \cite{GLMWY}, \cite{LMW}, \cite{Mallo}, \cite{Moraal}, \cite{Soder}.
We recall here some useful facts, following \cite{GLMWY}.

\smallskip

A Descartes configuration in $D$ dimensions consists of $D+2$
mutually tangent $(D-1)$-dimensional (hyper)spheres. We write
$S^{D-1}_a$ for a sphere of radius $a$. The curvature $c=1/a$ 
is endowed with positive sign for the orientation of $S^{D-1}_a$ 
with an outward pointing normal vector and negative for the opposite 
orientation. The curvatures of the spheres in a Descartes configuration
satisfy the quadratic Soddy--Gosset relation
\begin{equation}\label{Soddy}
 \left( \sum_{k=1}^{D+2} \frac{1}{a_k} \right)^2 = D \sum_{k=1}^{D+2} \left( \frac{1}{a_k} \right)^2 . 
\end{equation}
This relation can be formulated in matrix terms as $c^t Q_D c=0$, with $c=(1/a_1,\ldots, 1/a_{D+2})$
the vector of curvatures, and $Q_n$ the quadratic form determined by the matrix
$$ Q_D = I_{D+2} - D^{-1}\,\, 1_{D+2} 1_{D+2}^t, $$
where $1_{D+2}^t=(1,1,\ldots,1)$ and $I_{D+2}$ is the identity matrix. The augmented
curvature-center coordinates of a sphere $S^{D-1}_a$ with center $x=(x_1,\ldots,x_D)$ in 
$\R^D$ consist of a $(D+2)$-vector
$$ w =(\frac{\|x\|^2-a^2}{a},\frac{1}{a}, \frac{1}{a} x_1,\ldots, \frac{1}{a} x_D), $$
where the first coordinate describes the curvature of the sphere obtained from the given one
by inversion in the unit sphere. The reason for the first coordinate is so that one can extend 
unambiguously the augmented curvature-center coordinates to include the special case of degenerate 
spheres with zero curvature (hyperplanes). Given a Descartes configuration of spheres,
one assigns to it a $(D+2)\times (D+2)$ matrix $\cW$ whose $j$-th row is the vector of
augmented curvature-center coordinates of the $j$-th sphere in the configuration.
The space $\cM_D$ of all possible Descartes configuration in $D$ dimensions is then identified
with the space of all solutions $\cW$ to the equation
\begin{equation}\label{DescartesSol}
\cW^t \, Q_D \, \cW = \left(\begin{matrix} 0 & -4 & 0 \\ -4 & 0 & 0 \\ 0 & 0 & 2\, I_D
\end{matrix}\right).
\end{equation}
The space of solutions $\cM_D$ is endowed with a left and a right action of the
Lorentz group $O(D+1,1)$.

\smallskip

The $D$-dimensional Apollonian group $\cG_D$ is the group generated by the 
$(D+2)\times (D+2)$ matrices $S_j$ of the form
$$ S_j = I_{D+2} + \frac{2}{D-1} e_j 1^t_{D+2} - \frac{2D}{D-1} e_j e_j^t, $$
with $I_{D+2}$ the identity, $e_j$ the $j$-th standard coordinate vector, and
$1_{D+2}$ the vector with all coordinates equal to one. 

\smallskip

It is shown in \cite{GLMWY} that in dimension $D\geq 4$ the Apollonian group
$\cG_D$ is no longer a discrete subgroup of $\GL(D+2,\R)$ and its orbits on 
$\cM_D$ no longer correspond to sphere packings. However, the {\em dual Apollonian group} 
$\cG_D^\perp$ is a discrete subgroup of $\GL(D+2,\R)$, and the Apollonian packings of 
$(D-1)$-dimensional spheres we will be considering here are obtained, as in Theorem 4.3 of
\cite{GLMWY}, as orbits of the dual Apollonian group on $\cM_D$. The dual Apollonian
group $\cG_D^\perp$ is generated by reflections $S_j^\perp$ of the form
\begin{equation}\label{Reflex}
S_j^\perp =I_{D+2} + 2 \, 1_{D+2} e_j^t - 4 \, e_j e_j^t, 
\end{equation}
with $e_j$ the $j$-th unit coordinate vector. The matrix $S_j^\perp$ implements
inversion with respect to the $j$-th sphere. The Apollonian packing is obtained by
iteratively adding new Descartes configurations of spheres obtained from an initial
one by iteratively applying inversions with respect to some of the spheres. 
When $D\neq 3$ the only relations in the dual Apollonian group $\cG_D^\perp$
are $(S_j^\perp)^2=1$. Thus,
the spheres added at the $n$-th iterative step of the construction of the Apollonian
packing are in correspondence with all the possible reduced sequences 
$$ S_{j_1}^\perp S_{j_2}^\perp \cdots S_{j_n}^\perp, \ \ \  j_k \neq j_{k+1},\, \forall k, $$
acting on the point $\cW\in \cM_D$ that corresponds to the initial Descartes configuration. 
Clearly, there are $(D+2) (D+1)^{n-1}$ such sequences, hence the number of
spheres in the $n$-th level of the iterative construction is
$$ N_n :=\# \{ S^{D-1}_{a_{n,k}}\,:\, \text{ fixed } n \} =  (D+2) (D+1)^{n-1}. $$
In the following, we will focus on the case $D=4$, of Apollonian packings of $3$-spheres.

\medskip
\subsection{Lengths, packing constant, and zeta function}\label{zetaLSec}

We proceed as in \cite{CIL}, \cite{CIS} to associate a spectral triple to an
Apollonian packing $\cP_D$ of $(D-1)$-spheres in a $D$-dimensional space.
As above, let 
\begin{equation}\label{Lista}
\cL_D(\cP_D) = \{ a_{n,k}, \, n\in \N, \, 1\leq k \leq (D+2)(D+1)^{n-1} \} 
\end{equation}
be the list (with multiplicities) of the radii $a_{n,k}$ of the $(D+2)(D+1)^{n-1}$ spheres $S^{D-1}_{a_{n,k}}$
that are added in the $n$-th stage of the iterative construction of the packing.

The {\em packing constant} (or exponent of the packing), $\sigma_D(\cP_D)$ of
a packing $\cP_D$ of $(D-1)$-spheres is defined
as the exponent of convergence of the series
$$ \sum_{n\in \N} \sum_{k=1}^{(D+2)(D+1)^{n-1}} a_{n,k}^s, $$
that is,
\begin{equation}\label{sigmaD}
\begin{array}{rl}
\sigma_D(\cP_D) = & \displaystyle{\sup\{ s\in \R^*_+\,:\, \sum_{n\in \N} \sum_{k=1}^{(D+2)(D+1)^{n-1}} a_{n,k}^s =\infty \}} \\[3mm] = 
&\displaystyle{ \inf \{ s\in \R^*_+\,:\, \sum_{n\in \N} \sum_{k=1}^{(D+2)(D+1)^{n-1}} a_{n,k}^s<\infty \}}. \end{array}
\end{equation}
For $s> \sigma_D(\cP_D)$, one defines the zeta function $\zeta_{\cL_D}(s)$ as the sum
of the series
\begin{equation}\label{zetaLD}
 \zeta_{\cL_D}(s) = \sum_{n\in \N} \sum_{k=1}^{(D+2)(D+1)^{n-1}} a_{n,k}^s \, .
\end{equation} 

The zeta functions $\zeta_{\cL_D}(s)$, like the more general zeta functions of fractal
strings considered in \cite{Lapidus}, need not in general have analytic continuation to
meromorphic function on the whole complex plane, but there are a {\em screen} $\cS$, namely
a curve of the form $S(t)+it$, with $S:\R \to (-\infty, \sigma_D(\cP_D)]$, and a {\em window}
$\cW$ consisting of the region to the right of the screen curve $\cS$ in the complex plane, where
$\zeta_{\cL_D}(s)$ has analytic continuation. We refer the reader to \cite{Lapidus} for a
more detailed account of screens and windows for zeta functions of fractal strings.

\medskip
\subsection{Packing constant and Hausdorff dimension}\label{dimHSec}

The residual set of an Apollonian circle packing consists of the complement of
the union of all the open balls consisting of the interiors of the circles in the packing.
It was shown in \cite{Boyd} that the packing constant $\sigma_2$, defined as in
\eqref{sigmaD} is equal to the Hausdorff dimension of the residual set of the
circle packing. In the higher dimensional setting the problem of estimating the
Hausdorff dimension of the residual set of a packing of $(D-1)$-dimensional
spheres is much more involved, but there are some general estimates, obtained in
\cite{Larman} and \cite{Hawk}.

Consider the infimum of the packing constants
over all packings $\cP_D$,
$$\sigma_D=\inf_{\cP_D} \sigma_D(\cP_D).$$ 
Assuming all the spheres $S^{D-1}_{a_{n,k}}$ in the packing are
contained in the unit ball $B^D$, and denoting by $B^D_{a_{n,k}}$ the $D$-dimensional
ball with $\partial B^D_{a_{n,k}}=S^{D-1}_{a_{n,k}}$, the residual set of the packing is given by
$$ \cR(\cP_D)=B^D\smallsetminus \cup_{n,k} B^D_{a_{n,k}}. $$ 
Let $\dim_H(\cR(\cP_D))$ denote the Hausdorff dimension of the residual set and
$$\delta_D=\inf_{\cP_D} \dim_H(\cR(\cP_D))$$ the infimum over all packings of the Hausdorff
dimensions. The upper entropy dimension $h^+(\cR(\cP_D))$ of the residual set $\cR(\cP_D)$
is defined as 
$$ h^+(\cR(\cP_D))=\limsup_{\epsilon\to 0} - \frac{\log N_\epsilon(\cR(\cP_D))}{\log \epsilon}, $$
where for a set $X$, the number $N_\epsilon(X)$ counts the smallest number of sets of diameter
less than $2\epsilon$ that cover $X$. The lower entropy dimension is defined similarly, with a liminf
instead of limsup. It is known that the entropy dimension provides an upper bound for the
Hausdorff dimension. Then we have the following estimates (\cite{Larman} and \cite{Hawk}).

\begin{prop}\label{dimHsigma}
The radii $a_{n,k}$ of a packing $\cP_D$ satisfy $\sum_{n,k} a_{n,k}^D=1$ and
$\sum_{n,k} a_{n,k}^{D-1}=\infty$, hence $D-1< \sigma_D(\cP_D)\leq D$. 
The infima satisy $\delta_D\leq \sigma_D$, and for individual packings 
$\dim_H(\cR(\cP_D))\leq h^+(\cR(\cP_D))=\sigma_D(\cP_D)$.
\end{prop}

The identity $\sum_{n,k} a_{n,k}^D=1$ follows from the packing property, namely the
requirement that the residual set $\cR(\cP_D)$ in the $D$-dimensional unit ball has
zero $D$-dimensional volume. 
The value $\dim_H(\cP_D)$ is not known exactly. Some estimates are obtained,
with various methods, in \cite{Farr}, \cite{Moraal}, \cite{Soder}. We provide a simple
rough estimate in \S \ref{DimestSec} below, for the specific case of $3$-spheres.

\medskip
\subsection{Dimension estimate}\label{DimestSec}

Let $\cP=\cP_4$ be an Apollonian packing of $3$-dimensional spheres
$S^3_{a_{n,k}}$. We compute here a rough approximation to the 
packing constant $\sigma_4(\cP)$ of the Apollonian packing, defined
as in \eqref{sigmaD}.

\begin{prop}\label{approxsigma}
By replacing the collection of radii $\{ a_{n,k} \}$ in the 
$n$-th level of the Apollonian packing $\cP$ of $3$-spheres
with a single value $a_n =N_n/\gamma_n$, where $\gamma_n/N_n$ is
the average curvature in the $n$-th level, one obtains an approximate
estimate of the packing constant, 
$$ \sigma_{4,av}(\cP) \sim 3.85193\ldots $$
\end{prop}

\proof
As discussed above, the number of $3$-spheres in the 
$n$-th level of the packing $\cP$
is given by the number of reduced sequences in the generators of the group $\cG_n$,
namely
\begin{equation}\label{Leveln}
N_n:= \# \{ S^3_{a_{n,k}}\,:\, \text{ fixed } n \} = (D+2) (D+1)^{n-1} |_{D=4} = 6 \cdot 5^{n-1}.
\end{equation} 
Let $\gamma_n$ denote the sum of the curvatures of the spheres in the $n$-th level,
\begin{equation}\label{gamman}
\gamma_n =\sum_{k=1}^{6\cdot 5^{n-1}} \frac{1}{a_{n,k}}.
\end{equation}
As shown in Theorem 2 of \cite{Mallo}, the generating function of the 
$\gamma_n=\gamma_n(s)$ is
\begin{equation}\label{Gs}
G_{D=4}(u)= \frac{(1-x)(1-4x) u}{1-\frac{22}{3}x - 5 x^2},
\end{equation}
where $u=\gamma_0$ is the sum of the curvatures of the $D+2=6$ spheres 
in a Descartes configuration that gives the level-zero seed of the recursive 
construction.
We obtain from this an estimate of the metric dimension by replacing the
curvatures $1/a_{n,k}$ with their averages over levels. We denote the
resulting approximation to the dimension by $\sigma_{4,av}(\cP)$.
This is given by
$$ 
\sigma_{4,av}(\cP) = \lim_{n\to \infty} \frac{\log( 6\cdot 5^{n-1})}{\log\left( \frac{\gamma_n}{6\cdot 5^{n-1}}\right)}. 
$$
We expand \eqref{Gs} in a power series. Since the specific value of $u$ does not
influence the large $n$ behavior in the limit above, we look at the values for 
$u=1$, and we obtain
$$ G_{D=4} = \sum_{n=1}^\infty \gamma_n(1)\, x^n, $$
$$ \gamma_n(1) = \frac{(11+\sqrt{166})^n (-64+9\sqrt{166}) + (11-\sqrt{166})^n (64 +9 \sqrt{166})}{3^n \cdot 10\cdot \sqrt{166}}. $$ 
This then gives $\sigma_{4,av}(\cP) \sim 3.85193\ldots$ as stated.
\endproof

\medskip
\subsection{A spectral triple on the Cayley graph of the dual Apollonian group}\label{S3sec1}

Let $\cT_D$ denote the Cayley graph of the dual Apollonian group $\cG_D^\perp$.
Since for $D\neq 3$ the group $\cG_D^\perp$ is generated by the $D+2$ reflections
$S_j^\perp$ of \eqref{Reflex}, with the only relations of the form $(S_j^\perp)^2=1$,
the Cayley graph $\cT_D$ is an infinite tree with all vertices of valence $D+2$. We 
endow the tree $\cT_D$ with the structure of a {\em finitely summable tree},
in the sense of \S 7 of \cite{CIL}, by choosing a base vertex $v_0$ and endowing
all the $N_n=(D+2)(D+1)^{n-1}$ edges at a distance of $n$ steps from $v_0$ with
lengths $\ell(e_{n,k})=a_{n,k}$, equal to the radii of the spheres in the $n$-th level of the sphere
packing. Then, as in Theorem 7.10 of \cite{CIL} one obtains a finitely summable
spectral triple 
$$ ST_{\cT_D}=(\cA_{\cT_D},\cH_{\cT_D}, \cD_{\cT_D})
=\oplus_{e\in E(\cT_D)} (\cA_{\cT_D},\cH_{\ell(e)}, D_{\ell(e)}+\frac{\pi}{2\ell(e)} I). $$
The involutive subalgebra $\cA_{\cT_D}$ of the $C^*$-algebra $C(\cT_D)$ is determined,
as in \cite{CIL}, by the condition that $f \in \cA_{\cT_D}$ has $[\cD_{\cT_D},\pi(f)]$
densely defined and bounded, where $\pi: C(\cT_D) \to \cB(\cH_{\cT_D})$ is the
representation by bounded operators on the Hilbert space of the triple.
The pairs $(\cH_{\ell(e)}, D_{\ell(e)})$ are constructed as in the ``interval spectral
triple" of \S 3 of \cite{CIL}, with $\cH_\alpha=L^2([-\alpha,\alpha],\mu)$ with the
normalized Lebesgue measure $\mu$ and $D_\alpha$ with eigenvectors the 
basis elements $e_m =\exp(i\pi m x/\alpha)$ with eigenvalue $\pi m/\alpha$.
The Dirac operator $\cD_{\cT_D}$ then has spectrum 
$$ \Spec(\cD_{\cT_D})=\{ \frac{\pi(2m+1)}{2 \ell(e)} \,:\, e\in E(\cT_D),\, m\in \Z_+\} $$ $$ =
\{ \frac{\pi(2m+1)}{2 a_{n,k}} \,:\, n\in \N, \, 1\leq k\leq  (D+2)(D+1)^{n-1},\, m\in \Z_+\}. $$
The shift $\pi/2\ell(e)\, I$ to the Dirac operator $D_{\ell(e)}$ is introduced in \cite{CIL}
to avoid a kernel, so that the zeta function $\zeta_{D_{\cT_D}}(s)=\Tr(|\cD_{\cT_D}|^{-s})$ is
well defined.  
The zeta function of the Dirac operator of the spectral
triple $ST_{\cT_D}$ is given by
$$ \Tr(|\cD_{\cT_D}|^{-s})=\frac{2^{s+1}}{\pi^s}\, (1-2^{-s})\, \zeta(s) \,\, \zeta_{\cL_D}(s), $$
where $\zeta(s)$ is the Riemann zeta function, see \S 7.1 of \cite{CIL}.
The exponent of summability of the spectral triple (the metric dimension) is equal to the packing constant
of \eqref{sigmaD}, 
$$ \mathfrak{d}_{ST_{\cT_D}}= \sigma_D. $$

\medskip
\subsection{The spectral triple of a sphere packing}\label{S3sec}

Suppose given an Apollonian packing $\cP_D$ of $(D-1)$-dimensional
spheres $S^{D-1}_{a_{n,k}}$ in $\R^D$.
We modify the construction above, by introducing the contribution of the
$(D-1)$-spheres $S^{D-1}_{a_{n,k}}$ of the packing, through their respective spectral
triples. We replace the data $(\cH_{\ell(e_{n,k})}, D_{\ell(e_{n,k})})$ of the construction
above, for an edge $e_{n,k}$ of length $\ell(e_{n,k})=a_{n,k}$, with new data of the form
$(\cH_{S^{D-1}_{a_{n,k}}}, \cD_{S^{D-1}_{a_{n,k}}})$, where $\cH_{S^{D-1}_{a_{n,k}}}=
L^2(S^{D-1}_{a_{n,k}},\bS)$ is the Hilbert space of square integrable spinors on
the $(D-1)$-sphere $S^{D-1}_{a_{n,k}}$, and $\cD_{S^{D-1}_{a_{n,k}}}$ is the 
Dirac operator, with spectrum
$$ 
\Spec(D_{S^{D-1}_{a_{n,k}}})= \{ \lambda_{\ell,\pm}=\pm \, a_{n,k}^{-1}\, (\frac{D-1}{2} + \ell)\,:\, \ell\in \Z_+ \}
$$
and multiplicities
$$ {\rm Mult}(\lambda_{\ell,\pm}) = 2^{[ \frac{D-1}{2} ]} \binom{\ell+D}{\ell}. $$

\smallskip

\begin{defn}\label{S3defPD}
The spectral triple of the Apollonian packing 
$$ \cP_D=\{ S^{D-1}_{a_{n,k}} \,:\, n\in \N, \, 1\leq k \leq (D+2)(D+1)^{n-1} \}, $$
is given by 
\begin{equation}\label{PDSp3}
 (   \cA_{\cP_D}, \cH_{\cP_D}, \cD_{\cP_D}) = \oplus_{e\in \cE(\cT_D)} (\cA_{\cP_D}, 
 \cH_{S^{D-1}_{\ell(e)}}, \cD_{S^{D-1}_{\ell(e)}}),
\end{equation}
where $\cT_D$ is the Cayley graph of $\cG_D^\perp$, as above, with
edge lengths $\ell(e_{n,k})=a_{n,k}$, and the data
$(\cH_{S^{D-1}_{a_{n,k}}}, \cD_{S^{D-1}_{a_{n,k}}})$ are defined as above.
The involutive subalgebra $\cA_{\cP_D}$ consists of $f\in C(\cP_D)$ 
with $[\cD_{\cP_D},\pi(f)]$ densely defined and bounded.
\end{defn}

The fact that this is indeed a spectral triple follows from the 
general results of \cite{CIL} and \cite{CIS}. In particular, the
spectral action of the Swiss Cheese Cosmology model is
obtained by considering the case of a packing of $3$-dimensional
spheres,
\begin{equation}\label{P3S3}
{\rm ST}_{PSC} := (\cA_{\cP_4}, \cH_{\cP_4}, \cD_{\cP_4}).
\end{equation}

\smallskip

In order to compute the spectral action for the spectral triple of
a packing of $3$-spheres, we first recall some facts about the
spectral action of a single $3$-sphere.

\medskip
\subsection{The spectral action on the $3$-sphere}\label{S3Sp3sec}

We start by recalling some very simple and well known facts 
about the round sphere $S^3$ and its spectral action functional. We will need these
in the rest of this section as building blocks to construct the spectral triple 
and the spectral action for the Packed Swiss Cheese Cosmology.

The Dirac operator on the $3$-sphere $S^3$ with the round metric of unit radius
has spectrum $\Spec(D_{S^3})=\{ n+\frac{1}{2} \}$ with spectral multiplicities
${\rm Mult}(n+\frac{1}{2})= n(n+1)$, hence the spectral action 
takes the form
\begin{equation}\label{SactS3}
 \cS_{S^3}(\Lambda) = \text{Tr}(f(D_{S^3}/\Lambda)) = \sum_{n \in \mathbb{Z}}n(n+1)f((n+\frac{1}{2})/\Lambda). 
\end{equation}

\begin{lem}\label{zetaS3lem}
The zeta function of the Dirac operator is of the form
\begin{equation}\label{zetaDS3}
\zeta_{D_{S^3}}(s)=2\zeta(s-2,\frac{3}{2}) - \frac{1}{2} \zeta(s,\frac{3}{2}),
\end{equation}
where $\zeta(s, q)$ is the Hurwitz zeta function. The spectral triple $ST_{S^3}$
has simple dimension spectrum, with $\Sigma^+_{ST_{S^3}}=\{ 1,3 \}$. The
asymptotic expansion of the spectral action is correspondingly of the form
\begin{equation}\label{SAS3exp}
\cS_{S^3}(\Lambda) \sim \Lambda^3 f_3 - \frac{1}{4} \Lambda f_1
\end{equation}
\end{lem}

\proof The result immediately follows by writing
$$ \Tr(|D_{S^3}|^{-s})= \sum_{k\geq 0} 2(k+1)(k+2) \, (k+\frac{3}{2})^{-s} =
\sum_{k\geq 0} 2(k+\frac{3}{2})^{-(s-2)} -\frac{1}{2} \sum_{k\geq 0}  (k+\frac{3}{2})^{-s}. $$
The Hurwitz zeta function $\zeta(s, q)$ has a simple pole at $s=1$ with residue one,
hence $\zeta_{D_{S^3}}(s)$ has simple poles at $s=1$ and $s=3$, respectively with
residues ${\rm Res}_{s=1}\zeta_{D_{S^3}}(s)=-1/2$ and ${\rm Res}_{s=3}\zeta_{D_{S^3}}(s)=2$. Then applying \eqref{SAexpand}, one obtains the spectral action expansion. In the
constant term we have $\zeta_{D_{S^3}}(0)=2 \zeta(-2,3/2)- \zeta(0,3/2)/2=0$, 
since $\zeta(-2,3/2)=-1/4$ and $\zeta(0,3/2)=-1$.
\endproof

\begin{cor}\label{zetaS3a}
In the case of a $3$-sphere $S^3_a$ with the round metric of radius $a>0$,
the zeta function is of the form
\begin{equation}\label{zetaDS3a}
\zeta_{D_{S^3_a}}(s)=a^s (2\zeta(s-2,\frac{3}{2}) - \frac{1}{2} \zeta(s,\frac{3}{2})),
\end{equation}
and the asymptotic expansion of the spectral action is given by
\begin{equation}\label{SAS3expa}
\cS_{S^3_a}(\Lambda) \sim (\Lambda a)^3 f_3 - \frac{1}{4} (\Lambda a) f_1.
\end{equation}
\end{cor}

\proof The spectrum 
of the Dirac operator $D_{S^3_a}$ is a scaled copy $\frac{1}{a}(\frac{1}{2}+\mathbb{Z})$
of the spectrum of $D_{S^3_1}$, and the multiplicities coincide.
Thus, we have
$$ \Tr(|D_{S^3_a}|^{-s}) = \sum_{n=1}^\infty 2n(n+1)\left(\frac{n+1/2}{a}\right)^{-s} = 2a^s 
\sum_{n=1}^\infty n(n+1)(n+1/2)^{-s}$$
$$ 
= 2a^s \sum_{n=1}^\infty (n+1/2)^2(n+1/2)^{-s} - \frac{a^s}{2}\sum_{n=1}^\infty (n+1/2)^{-s}$$
$$ 
= 2a^s\sum_{n=0}^\infty (n+3/2)^{-(s-2)} - \frac{a^s}{2}\sum_{n=0}^\infty (n+3/2)^{-s}$$
When $\Re(s)>3$ (the metric dimension of the 3-sphere), this simplifies to \eqref{zetaDS3a}.
\endproof

\smallskip

A method for non-perturbative computations of  the spectral action
functional based on the Poisson summation formula
was developed in \cite{CC-uncanny},
for sufficiently regular geometries for which the
Dirac spectrum and the spectral multiplicities are
explicitly known. In particular, the
spectral action for the round sphere $S^3$ was computed 
in \cite{CC-uncanny} using this method. The computation was
generalized to spherical space forms, \cite{Teh},
and to $3$-dimensional tori and Bieberbach manifolds in
\cite{MaPieTeh}, \cite{MaPieTeh2}, \cite{OlSi}.
The computation of \cite{CC-uncanny} for the $3$-sphere can be
summarized quickly as follows.
Let $f$ be a rapidly decaying even function. The eigenvalues of $D_{S^3}$ 
form an arithmetic progression, and there is a polynomial 
$P(u)= u^2 -\frac{1}{4}$ that interpolates the spectral multiplicities, 
$\text{Mult}(\lambda)=P(\lambda)$. Thus, one can write the spectral action as
$$ \cS_{S^3}(\Lambda) = \sum_{n \in \mathbb{Z}} g(n+\frac{1}{2}), $$
where  $g(u) = (u^2 - \frac{1}{4})f(u/\Lambda)$ is also a rapidly decaying function. 
This is then the sum of values of a rapidly decaying function on points of a lattice,
which can be evaluated using the Poisson summation formula
$$\sum_{n \in \mathbb{Z}}g(n+\frac{1}{2}) = \sum_{n \in \mathbb{Z}}(-1)^n\hat{g}(n) ,$$
where 
$$\hat{g}(x) = \int_\mathbb{R} g(u)e^{-2\pi ixu}du = \int_\mathbb{R} (u^2-\frac{1}{4}) f(u/\Lambda) e^{-2\pi ixu} du$$
is the Fourier transform 
$$\hat{g}(x) = \Lambda^3 \int_\mathbb{R}v^2f(v)e^{-2\pi i\Lambda xv}dv - \frac{1}{4}\Lambda\int_\mathbb{R}f(v)e^{-2\pi i\Lambda xv}dv,$$
after substituting $u = \Lambda v$. Let $\hat{f}^{(2)}$ denote the Fourier transform of $v^2f(v)$, in the first term above. It is shown in \cite{CC-uncanny} that 
the sum on the Fourier transformed side can be very accurately
approximated by the term with $n=0$, yielding for any $k$
$$\text{Tr}(f(D/\Lambda)) = \Lambda^3 \int_\mathbb{R}v^2f(v)dv - 
\frac{1}{4} \Lambda \int_\mathbb{R}f(v)dv + O(\Lambda^{-k}). $$
In the case of the round $3$-sphere $S^3_a$ of radius $a$, we have 
$$\Tr(f(\frac{D_{S^3_a}}{\Lambda})) = \Tr(f(\frac{D_{S^3_1}}{\Lambda a})), $$
and the approximation formula above extends to $S^3_a$, replacing $\Lambda$ 
with $\Lambda a$, so one obtains 
\begin{equation}\label{SAS3a}
\text{Tr}(f(D_{S^3_a}/\Lambda)) = (\Lambda a)^3 \int_\mathbb{R}v^2f(v)dv - 
\frac{1}{4} (\Lambda a) \int_\mathbb{R}f(v)dv + O((\Lambda a)^{-K}),
\end{equation}
for arbitrary $K\in \N$, 
which agrees with the expression \eqref{SAS3expa}, with the error term
as in \cite{CC-uncanny}.

\medskip
\subsection{Zeta function of a $3$-sphere packing}\label{ZetaSpActSec}

We focus here on the case of a packing $\cP=\cP_4$ of $3$-spheres, 
where at the $n$-th iterative
step in the construction one has $6\cdot 5^{n-1}$ spheres, with radii $a_{n,k}$ with
$k=1,\ldots, 6\cdot 5^{n-1}$, starting with an initial Descartes configuration of 
$6$ mutually tangent  $3$-spheres. As above, 
let $\cL =\cL_4= \{ a_{n,k}\,|\, n\in \N, \, k\in\{ 1,\ldots, 6\cdot 5^{n-1} \} \}$ be
the length spectrum of the radii of all the $3$-spheres in the packing. We consider
the associated zeta function \eqref{zetaLD} for $D=4$, which we denote simply 
by $\zeta_\cL(s)$,
\begin{equation}\label{zetaL}
\zeta_\cL(s) := \sum_{n\in \N} \sum_{k=1}^{6\cdot 5^{n-1}} a_{n,k}^s. 
\end{equation}

\begin{prop}\label{zetaST} Let $\sigma_4(\cP)$ be the packing constant of 
an Apollonian packing $\cP$ of $3$-dimensional spheres, as in \eqref{sigmaD}.
For $s> \sigma_4(\cP)$, the zeta function of the Dirac operator $\cD_\cP$ 
of the spectral triple ${\rm ST}_{PSC}$ of \eqref{P3S3} is given by
\begin{equation}\label{zetaPSCC}
 \Tr(|\cD_\cP|^{-s}) =\left( 2\zeta(s-2, \frac{3}{2}) - \frac{1}{2}\zeta(s, \frac{3}{2}) \right) 
 \zeta_\cL(s),
\end{equation}
where $\zeta(s, q)$ is the Hurwitz zeta function and $\zeta_\cL (s)$ is as in \eqref{zetaL}.
\end{prop}

\proof
Since $0 \notin \text{Spec}(D_{S^3_a})$, $D_{S^3_a}$ is invertible and so
is then the Dirac operator $\cD_\cP$ for the spectral triple ${\rm ST}_{PSC}$.
The metric dimension is then given by $\inf \{ \beta > 0 \mid \Tr (|\cD_\cP|^{-\beta}) < \infty \}$,
where the zeta function is given by 
$$ \Tr(|\cD_\cP|^{-s}) = \sum_{n=1}^\infty \sum_{k=1}^{6\cdot 5^{n-1}} 
\Tr(|D_{S^3_{a_{n,k}}}|^{-s}). $$
Each term in this sum can be computed as in \eqref{zetaDS3a}.
We can then evaluate the zeta function of the spectral triple ${\rm ST}_{PSC}$, using the
fact that the contribution of each sphere $S^3_{a_{n,k}}$ is of the form
$\Tr (|D_{S^3_{a_{n,k}}}|^{-s}) = a_{n,k}^s ( 2\zeta(s-2, \frac{3}{2}) - \frac{1}{2}\zeta(s, \frac{3}{2}) )$, 
and we obtain 
$$ \Tr(|D|^{-s}) = \sum_{k=0}^\infty \text{Tr}(|D_{S^3_{a_{n,k}}}|^{-s}) 
= \left( 2\zeta(s-2, \frac{3}{2}) - \frac{1}{2}\zeta(s, \frac{3}{2}) \right) \sum_{k=0}^\infty a_{n,k}^s , $$
for $s> \sigma$, with $\sigma=\max\{3,1,\sigma_4(\cP) \}=\sigma_4(\cP)$, where 
$\sigma_D(\cP_D)$, as in \eqref{sigmaD}, is the packing constant of $\cP_D$, the 
exponent of convergence of the series $\sum a_{n,k}^s$. We know from \S \ref{dimHSec}
that $3\leq \sigma_4 \leq 4$, hence $\max\{3,1,\sigma_4(\cP) \}=\sigma_4(\cP)$.
\endproof

\smallskip
\subsection{Dimension spectrum}\label{DimSpSec}

The definition of dimension spectrum we are using in this paper is
the same as in \cite{CoMa-book}. It is slightly different from other versions
in the literature, see \cite{CoMo} and \cite{IoLe}. In particular, note that
the dimension spectrum $\Sigma_M$ for an ordinary smooth
manifold $M$ of dimension $n=\dim M$ the dimension spectrum is given by the set
$\Sigma_M =\Sigma^+_M=\{ 0, 1, 2, \cdots, n \}$, according to Example 13.8 of \cite{CoMa},
or by $\Sigma_M=\{ m \in \Z\,:\, m\leq n \}$, according to \cite{Co-qgr}, p.22, and 
Proposition A.2 of \cite{IoLe}. The leading terms in the asymptotic expansion
of the spectral action, which correspond to the gravitational terms in the
action functional, arise from the points in $\Sigma^+_M=\Sigma_M\cap \R_+$,
which are the same in all cases, hence for our purposes the slight discrepancy 
between different versions of the notion of dimension spectrum adopted in
the literature does not affect the results.

\smallskip

In the following, we will focus on analyzing the poles in $\R^*_+$ and off
the real line of the zeta function $\zeta_{\cD_\cP}(s)= \Tr(|\cD_\cP|^{-s})$.
While these poles certainly contribute points to the dimension spectrum,
there may, in principle, be additional poles coming from other zeta functions
$\zeta_{b,\cD_\cP}(s)= \Tr( b |\cD_\cP|^{-s})$, for algebra elements $b\in \cB$
not equal to the identity.  In the case of smooth manifolds, it is known
(see for instance Proposition A.2 of \cite{IoLe}) that these zeta functions
do not contribute additional poles. While there is no general result for
arbitrary spectral triple, in the case of the spectral triple of a fractal geometry
it is often suggested that the subalgebra of ``smooth functions" should
consist of functions that are supported on finitely many levels of the
fractal construction (for example, in the case of a Cantor set, that would
mean locally constant functions). If the fractal is built out of pieces that
are smooth manifolds (as in the case of a sphere packing) then one
should also require that the functions are smooth on each smooth 
component. While this choice of smooth subalgebra does not necessarily
have, in general, the same good analytic properties as the algebra of 
smooth functions on a smooth manifold, it is a natural choice in this
setting. In this case, the fact that the additional zeta functions 
$\zeta_{b,\cD_\cP}(s)$ do not contribute new poles can then be
reduced to the known case of manifolds. When we discuss perturbations
of the Dirac operator by a scalar field, to obtain a slow-roll potential for
inflationary models, we will assume that the scalar fields also
live in this smooth subalgebra. 

\smallskip

The result of Proposition \ref{zetaST} then shows that the
dimension spectrum of the spectral triple ${\rm ST}_{PSC}$
is given by the following set.

\begin{lem}\label{DimSpPSC}
The dimension spectrum $\Sigma_{PSC}$ of the spectral triple ${\rm ST}_{PSC}$
consists of the union of the dimension spectrum of the $3$-sphere, a single other real
point $\sigma_4(\cP)$, and a countable collection of points off the real line, lying in
the window $\cW$ where $\zeta_{\cL}(s)$ has analytic continuation.
\end{lem}

In general it is difficult to characterize precisely the position of the
poles that are off the real line, except in the case of self-similar
fractals. We will discuss how to obtain some control on the
contributions of these points to the expansion of the spectral
action in the following section.

\section{Spectral Action for Packed Swiss Cheese Cosmology}\label{SAPSCsec}

In this section we use the results of the previous section on the
zeta function of the Dirac operator on the Packed Swiss Cheese Cosmology
in order to study how the spectral action is affected by the presence
of fractality. In particular, under some restrictive assumptions on the
analytic properties of the zeta function $\zeta_\cL(s)$ of the Apollonian
packing, and using the relation between the heat kernel
and the zeta function and results on the asymptotic expansion of
the heat kernel, we will obtain an expansion of the spectral action
that contains the familiar gravitational terms of a $3$-dimensional
sphere, but also has additional terms determined by the residue of
the zeta function at the packing constant, and a Fourier series
of additional oscillatory terms coming from fluctuations produced
by the presence of poles of the zeta function located off the real line. 

\smallskip
\subsection{Zeta function, heat kernel, and spectral action on fractals}\label{OscSec}

An asymptotic expansion for the spectral action, in the sense of \cite{CC}, is
known to exist (see Theorem 1.145 of \cite{CoMa-book}) whenever there is a 
small-time asymptotic expansion for the heat kernel of the corresponding 
Dirac operator.  In the case of an ordinary manifold, or an almost-commutative
geometry, the heat kernel expansion is known by classical results on 
pseudo-differential operators. For more general spaces, like fractal geometries,
there are no analogous theorems that hold with the same level of generality,
although several results on the heat kernel expansion on fractals are
available, see for instance the detailed survey given in \cite{Dunne}. 
For some general results about Laplacians on fractals and heat kernels 
we also refer the reader to \cite{Kigami}, \cite{Stri}.

Specifically in relation to the asymptotic expansion of the spectral action,
cases where the zeta function has poles off the real line, which contribute
log-oscillatory terms to the spectral action, were studied in \cite{EIS} and
\cite{EckZa}.

\smallskip

The main new feature that arises in the case of fractal geometries is,
as we have seen in the previous section, the presence of poles of
the zeta function that are off the real line. In the case of the geometry
of the Apollonian packings of $3$-spheres we consider in this paper,
those poles correspond to the poles off the real line of the zeta function 
$\zeta_\cL(s)$ of the length spectrum $\cL=\cL(\cP)$ of the packing. 

\smallskip

As discussed in \S 1--3 of \cite{Lapidus}, for general zeta functions of
fractal strings $\cL$ the distribution of the non-real poles can be very
complicated. In the best possible case, which corresponds to fractals 
with a self-similar structure where the contraction ratios are
all integer powers of a fixed scale $0<r<1$ (lattice case) the non-real
poles lie, periodically spaced, on finitely many vertical lines. In cases
with self-similar structure, but where the contraction ratios do not
satisfy the lattice condition (non-lattice case), the poles off the real
line have a quasi-periodic behavior and are approximated by a sequence
of lattice strings. 

\smallskip

In the case of a length spectrum with exact self-similarity realized by a single contraction
ratio $r$, the poles off the real line lie on the vertical line with $\Re(s)=\sigma$, which is
the Hausdorff dimension, and with periodic spacings of length $\frac{2\pi}{\log (1/r)}$,
namely $s= \sigma + \frac{2\pi i m}{\log (1/r)}$ with $m\in \Z$. We will discuss
in \S \ref{dodecaSec} an example of this kind, which is relevant to our
cosmological models. In such cases with exact self-similarity, it is known (see
\S 4 of \cite{Dunne}) that the contribution of the off-real poles to the heat-kernel
asymptotic consists of a series of log-oscillatory terms. We have the following model
case for this situation.

\begin{prop}\label{singlescale}
Let $X$ be a fractal geometry with a Dirac operator $D_X$ of the associated
spectral triple with the following property: the eigenvalues of $|D_X|$ 
grow exponentially like $b^n$, for some $b>1$, and the 
spectral multiplicities also grow exponentially like $a^n$ for some $a>1$. Then 
the spectral action $\cS_X(\Lambda)=\Tr(f(D_X/\Lambda))$ has an expansion for large
$\Lambda$ of the form
\begin{equation}\label{SAexp1fractal}
 \cS_X(\Lambda) \sim \Lambda^\sigma \sum_{m\in \Z} \Lambda^{\frac{2\pi i m}{\log b}} \,\,
f_{s_m}  
\end{equation}
where $s_m = \sigma + \frac{2\pi i m}{\log b}$ and $\sigma = \frac{\log a}{\log b}$.
The coefficients $f_{s_m}$ are given by integrals 
$$ f_{s_m} =\frac{1}{\log b} \int_0^\infty f(u) u^{s_m-1} du. $$
For sufficiently rapidly decaying test functions $f(u)$, 
the Fourier series $\sum_m \Lambda^{\frac{2\pi i m}{\log b}} f_{s_m}$ converges uniformly to
a smooth function $\ff_\sigma(\theta)$ of the circle variable $\theta =\frac{\log \Lambda}{\log b}$
modulo $2\pi \Z$. 
\end{prop}

\proof The zeta function has the form $\zeta_{D_X}(s) =\sum_n a^n b^{-sn} = (1-ab^{-s})^{-1}$,
with simple poles at $s=\frac{\log a}{\log b} + \frac{2\pi i m}{\log b}$, and with exponent of
convergence $\sigma=\frac{\log a}{\log b}$.
The trace of the heat kernel has the ``exponential form"
\begin{equation}\label{expformheat}
 \Tr( e^{- t D^2} ) = \sum_n a^n e^{-t b^{2n}} 
\end{equation} 
for some constants $a,b$. Then it is known (see \S 4.2 of \cite{Dunne})
that one has a small-time asymptotics of the form 
\begin{equation}\label{asymptfractal}
 \Tr( e^{- t D^2} ) \sim \frac{t^{-\frac{\log a}{2\log b}}}{2\log b}  \sum_{\m\in \Z} 
\Gamma(\frac{\log a}{2\log b}+\frac{\pi i m}{\log b}) \exp( -\frac{\pi i m}{\log b} \log t )
\end{equation}
$$ = \frac{1}{2 \log b}  \sum_m \Gamma(s_m/2)\,\, t^{-s_m/2}. $$
Indeed, through the Mellin transform relation between the heat
kernel and the zeta function 
$$ |D_X|^{-s} = \frac{1}{\Gamma(s/2)} \int_0^\infty e^{-t D_X^2} \,\, t^{\frac{s}{2}-1} \, dt, $$
this corresponds to
$$ \zeta_{D_X}(s)=\Tr(|D_X|^{-s}) =\sum_m \frac{\Gamma(s_m/2)}{\Gamma(s/2) \cdot (s-s_m) \cdot \log b } + {\rm holomorphic} $$ 
with poles at $s=s_m=\sigma +\frac{2\pi i m}{\log b}$ with residue $1/\log b$. To obtain then
an expansion for the spectral action, one proceeds as in Theorem 1.145 of \cite{CoMa-book}.
One considers a test function written as Laplace transform as $k(u)=\int_0^\infty e^{-x u} h(x) dx$,
so that $k(tD_X^2)=\int_0^\infty e^{-xt D_X^2} h(x) dx$. Using the expansion \eqref{asymptfractal}
one then has
$$ k(tD_X^2) \sim \sum_m \frac{\Gamma(s_m/2)}{2\log b}\,\, t^{-s_m/2} \int_0^\infty x^{-s_m/2} h(x) dx. $$
Since $\Re(s_m)=\sigma>0$, we can write $x^{-s_m/2}$ as Mellin transform
$$ x^{-s_m/2} = \frac{1}{\Gamma(s_m/2)} \int_0^\infty e^{-x v} \, v^{\frac{s_m}{2}-1} \, dv, $$
hence we obtain
$$ \Tr(k(tD_X^2)) \sim  \sum_m {\rm Res}_{s=s_m} \zeta_{D_X}(s) \,\, t^{-s_m/2} \,
\int_0^\infty k(v) v^{\frac{s_m}{2} -1} dv. $$
Then taking $f(u)=k(u^2)$ we obtain
$$ \int_0^\infty k(v) v^{\frac{s_m}{2} -1} dv = 2 \int_0^\infty f(u) u^{s_m-1} du. $$
We then set $t = \Lambda^{-2}$ to obtain the form of the spectral action and the 
expansion 
$$ \cS_X(\Lambda) \sim \Lambda^\sigma \sum_{m\in \Z}  \Lambda^{\frac{2\pi i m}{\log b}} \,\, 
(\int_0^\infty f(u) u^{s_m-1} du) \,\, {\rm Res}_{s=s_m} \zeta_{D_X}(s), $$
which gives \eqref{SAexp1fractal}. 
Using the relation between Mellin and Fourier transform,
we can rewrite the coefficients
$$ f_{s_m} = \frac{1}{\log b} \int_0^\infty f(u) u^{\sigma} \,\, e^{-2 \pi i m \frac{\log u}{\log b}} \,\, \frac{du}{u}
= \int_\R F(\lambda) e^{-2\pi i m \lambda} d\lambda =2\pi \hat F(-2\pi m), $$
where $\lambda = \frac{\log u}{\log b}$ and $F(\lambda)=f(b^\lambda) b^{\lambda \sigma}$,
and $\hat F(\xi)=(2\pi)^{-1} \int_\R F(\lambda) e^{i \xi \lambda} d\lambda$ is the Fourier transform.
Provided the test function $f$ is sufficiently rapidly decaying, the function $F(\lambda)$ is also a
rapidly decaying function (at $\lambda \to +\infty$ because of the behavior of $f$ and at $\lambda \to -\infty$ because of the term $b^{\lambda \sigma}$). Then 
the Fourier transform $\hat F(\xi)$ is also rapidly decaying, hence the 
Fourier series $\sum_m \Lambda^{\frac{2\pi i m}{\log b}} f_{s_m}=\sum_m f_{s_m} e^{2\pi i m \theta}$ 
converges to a smooth function $\ff_\sigma(\theta)$. 
\endproof

\smallskip

More generally, in the case of exact self-similarity realized by a set of contraction
ratios $\{ r_1, \ldots, r_m \}$, the zeta function $\zeta_\cL(s)$ has a denominator
of the form $1-\sum_j r_j^{-s}$. The exponent of convergence is the self-similarity
dimension given by the real number $\sigma$ satisfying the self-similarity equation
$\sum_{j=1}^m r_j^{-\sigma}=1$. If the scaling factors $r_j$ satisfy the lattice condition,
namely if the subgroup 
$\prod_{j=1}^m r_j^\Z \subset \R^*_+$ is discrete, then (see Theorem 2.17 of \cite{Lapidus}) 
the complex poles lie on finitely many vertical lines with $\Re(s)\leq \sigma$, and are periodically
spaced with period $2\pi /\log(r^{-1})$, where $r$ is the multiplicative generator of the
scaling group, with $r_j = r^{n_j}$ for some integers $n_j$. In this lattice case, assuming all the
poles are simple and there are no cancellations from a numerator of $\zeta_\cL(s)$, one still
obtains an asymptotic expansion of the form \eqref{asymptfractal} with one log
oscillatory series for each of the finitely many vertical lines containing the complex poles
of $\zeta_\cL(s)$.

\smallskip

In the case with exact self-similarity realized by a set of contraction
ratios $\{ r_1, \ldots, r_m \}$ that do {\em not} satisfy the lattice condition, 
it is no longer true that the complex poles lie on finitely many vertical lines.
It is known (Theorem 2.17 of \cite{Lapidus}) that in this case there are no
other poles on the line $\Re(s)=\sigma$ except the real pole $s=\sigma$, but
there is a sequence of complex poles approaching the vertical line $\Re(s)=\sigma$
from the left. Moreover, all the complex poles are contained in a vertical strip 
$\sigma_0 \leq \Re(s)\leq \sigma$, for some $\sigma_0\in \R$. Moreover, in this
general non-lattice case, the complex poles can be approximated by the poles of
an infinite family of lattice cases, with increasingly large oscillation periods 
(see \S 3 of \cite{Lapidus}), which in turn correspond to an infinite family of 
Fourier series of log-oscillatory terms.

\smallskip

\begin{rem}\label{ApolloSelfSim}{\rm
In the case of the $D=2$ Apollonian circle packings, there are known results that
characterize the presence of self-similarity, \cite{ChiDo}: these packings 
correspond to quadratic irrationalities, via a continued fractions argument. However,
analogous results for the higher dimensional Apollonian packings, characterizing
the presence of exact self-similarity,s are not presently known. }
\end{rem}

\smallskip
\subsection{Approximations and expansion}\label{ApproxSec}

In more general situations, even for nice cases of fractal geometries
with exact self-similarity, we do not have such explicit control over
the oscillatory terms as in the case of Proposition \ref{singlescale},
where one has a single scale factor for self-similarity. In particular,
in cases of self-similar geometries that do not satisfy the lattice
conditions, the oscillatory terms can only be described via a sequence
of approximations. Thus, we need to introduce some choices of
approximations in the description of the log-oscillatory contributions
to the spectral action coming from the poles of the zeta function
that are off the real line.

\smallskip

A first, very rough approximation, which we will occasionally use in the
following, consists of replacing the smooth function $\ff_\sigma(\theta)$
in the expansion $\cS_X(\Lambda) \sim \Lambda^\sigma \, \ff_\sigma(\theta(\Lambda))$
of Proposition \ref{singlescale} with its average value on the circle. This
corresponds to selecting only the zero order Fourier coefficient
$$ \frac{1}{2\pi} \int_0^{2\pi} \ff_\sigma(\theta) d\theta = f_\sigma = 
\frac{1}{\log(b)} \int_0^\infty f(u) u^{\sigma -1} du = 
{\rm Res}_{s=\sigma} \zeta_{D_X}(s) \cdot  \int_0^\infty f(u) u^{\sigma -1} du. $$
This corresponds to only counting the contribution of the pole $s=\sigma$
on the real line and neglecting the contributions of the poles that lie off
the real line. 

\smallskip

In a similar way, one can decide to approximate the function
$\ff_\sigma(\theta(\Lambda))$ by truncating the Fourier
series at a higher order. In the case of a fractal geometry
with the non-lattice property,  where there is an infinite
sequence of lattice approximations (\S 3 of \cite{Lapidus})
to the off-real poles of the zeta function, these give rise
to terms with increasingly long oscillation periods in the
expansion of the spectral action. One can then choose to
truncate the Fourier series at some fixed size $M = m/\log b$,
so that oscillatory series with longer oscillation periods get
truncated earlier and contribute less to the approximation.

\smallskip

Note that truncating the Fourier series in the spectral action
expansion at some size $M = m/\log b$ can also be seen as 
truncating the series of oscillatory terms in the heat kernel
expansion \eqref{asymptfractal}. The size of these terms
is determined by the size of the value of the Gamma function
$\Gamma(\frac{\log a}{2\log b}+\frac{\pi i m}{\log b})$. The
Gamma function decays exponentially fast along the vertical
line $\Re(s)=\frac{\log a}{2\log b}$, hence these oscillatory
terms in the heat kernel expansion become rapidly very small
in comparison to the contribution of the $m=0$ term.

\smallskip

In all of these cases, when we introduce approximations to
the oscillatory terms, the approximation we obtain for the
spectral action is no longer really an ``asymptotic expansion" 
in the sense of \cite{Hardy}. The usual meaning of asymptotic
expansion implies that the function can be approximated
around some value of the argument (or a limit value) up to
arbitrary high order. For the purpose of building gravitational models, 
it will suffice to obtain an expansion of the spectral action up to
order $\Lambda^0$ (including the oscillatory terms), and some
sufficiently good approximation in cases where the oscillatory 
terms cannot be fully computed explicitly. For this reason, in
the following we will simply use the terminology ``expansion"
of the spectral action, rather than insisting on the stronger
properties of a genuine asymptotic expansion.

\smallskip
\subsection{Some analytic assumptions}\label{AnSec}

As we pointed out in Remark \ref{ApolloSelfSim}, 
unlike the Apollonian circles case, in dimension $D=4$ we do not have 
a characterization of the presence of exact self-similarity in the sphere
packing. However, in order to obtain a reasonably behaved model, with
respect to the properties of the zeta function and the spectral action
functional, we restrict our attention to a subset of all the possible 
Apollonian packing, identified by a set of requirements on the properties of
the associated zeta function $\zeta_\cL(s)$.

\begin{defn}\label{defan}{\rm 
A packing $\cP$ of $3$-dimensional spheres is {\em analytic} if 
it satisfies the following properties:
\begin{enumerate}
\item The zeta function $\zeta_\cL(s)$ of the packing $\cP$ has analytic
continuation to a meromorphic function on a region of the complex plane that contains 
the non-negative real axis.
\item The analytic continuation $\zeta_\cL(s)$ has only one pole on the non-negative
real axis, located at $s=\sigma_4(\cP)$.
\item The poles of $\zeta_\cL(s)$ are simple.
\item There is a family $\cL_n$, $n\in \N$, of self-similar fractal strings with the
lattice property, and with increasingly large periods, such that the complex poles of 
$\zeta_{\cL}(s)$ are approximated by the complex poles of $\zeta_{\cL_n}(s)$. 
\end{enumerate}}
\end{defn}

In terms of screens and windows, as in \cite{Lapidus}, the first condition above consists of 
the property that the screen function $S: \R \to (-\infty, \sigma_4(\cP)]$
satisfies $S(0)< 0$.

\smallskip

In the last condition, the period of a self-similar fractal string $\cL_n$ with the
lattice property is the length $\pi_n:=\frac{2\pi}{-\log r_n}$ with the property that all the
poles of $\zeta_{\cL_n}(s)$ off the real line lie on finitely may vertical lines $\Re(s)=\sigma_j$
with periodic spacing by $\frac{2\pi}{-\log r_n}$. The approximation condition means that,
for all $\epsilon >0$, there exists an $n\in \N$ an $R=R(\epsilon,n)>0$, such
that, within a vertical region of size at most $R$, the complex poles of $\zeta_{\cL}(s)$ 
are within distance $\epsilon$ of the poles of $\zeta_{\cL_n}(s)$. For more details
see \S 3.4.1 of \cite{Lapidus} and see Figure 3.6 of \cite{Lapidus} for an explicit
example of such an approximation.

\smallskip
\subsection{Heuristics of analytic assumptions}\label{heurSec}

At present, we do not have a
characterization of the locus of packings satisfying the constraints listed in Definition \ref{defan}
(for example, in terms of a geometric locus in the configuration space $\cM_D$
of Descartes configurations). 
We can, however, provide some heuristic explanation for the geometric meaning of the
requirement that the zeta function $\zeta_{\cL_D}(s)$ of the length spectrum 
$\cL_D=\{ a_{n,k} \}$ of an Apollonian packing $\cP_D$ of $(D-1)$-dimensional spheres
satisfies these properties.

\smallskip

Consider the possibility that a sphere packing has exact self-similarity. This would
mean that there is a finite set $\{ r_1, \ldots, r_m \}$ of scaling ratios, with the
property that, for all $n,k$, the radii $a_{n,k} \in \R^*_+$ of the packing belong to
the subgroup $\prod_{j=1}^m r_j^\Z \subset \R^*_+$. This subgroup will, in
general, be dense in $\R^*_+$ (non-lattice case). 
In such cases, for the zeta function of \eqref{zetaPSCC},
the factor $\zeta_\cL(s)$ would have analytic continuation to a meromorphic function
on all of $\C$ (see Theorem 2.4 of \cite{Lapidus}), hence the first condition of
Definition \ref{defan} would certainly be satisfied. Moreover, in such a case,
the second condition would be satisfied by Theorem 2.17 of \cite{Lapidus}.
The third condition would be satisfied, at least in the general case (again by
Theorem 2.17 of \cite{Lapidus}). The last condition is obvious in the lattice
case, and is a consequence of the approximation result of \S 3 of \cite{Lapidus}
in the non-lattice case. In the non-lattice self-similar case, the self-similar
strings $\cL_n$ with the lattice conditions are constructed using Diophantine
approximation (Lemma 3.16 and Theorem 3.18 of \cite{Lapidus}).

\smallskip

Thus, one should think of the conditions of Definition \ref{defan} as a 
generalization of the good conditions satisfied by the zeta function of
a fractal with exact self-similarity. One can expect that they may be
fulfilled by especially regular (especially symmetric) choices of
Descartes configuration, although we leave a more precise mathematical
investigation of this question to future work.

\smallskip
\subsection{Spectral action expansion}

For the rest of this section we make the assumptions that the packing $\cP$ of
$3$-dimensional spheres we are considering satisfies the conditions listed in Definition \ref{defan}.

\medskip

Under the assumptions of Definition \ref{defan}, the result of Lemma \ref{DimSpPSC}
on the dimension spectrum $\Sigma_{ST_{PSC}}$ can be refined to the following form.

\begin{prop}\label{propDimSp}
For a packing $\cP$ of $3$-spheres satisfying the properties of Definition \ref{defan}, 
the non-negative dimension spectrum of the spectral triple of the Packed Swiss
Cheese Cosmology consists of the points 
\begin{equation}\label{SigmaplusPSC}
\Sigma^+_{ST_{PSC}} = \{  1,3,  \sigma_4(\cP) \}, 
\end{equation}
with the  metric dimension of the spectral triple is $\mathfrak{d}_{PSC}= \sigma_4(\cP)$.
The spectral triple has simple dimension spectrum, which, in addition to the points of
$\Sigma^+_{ST_{PSC}}$ on the real line contains a countable family of points
off the real line, contained in a horizontally bounded strip 
$\sigma_{min} \leq \Re(s) \leq \sigma_4(\cP)$, approximated by the non-real poles
of a family $\cL_n$ of self-similar fractal strings with the lattice property and with
increasingly large oscillation periods. 
\end{prop}

\proof Under the assumptions that the zeta function $\zeta_\cL(s)$ has
analytic continuation to a window including the positive real axis, we see
that the zeta function $\zeta_{\cD_\cP}(s)=\Tr(|\cD_\cP|^{-s})$ of the Dirac 
operator of the spectral triple $ST_{PSC}$ also has
analytic continuation to a meromorphic function in the same region.
Moreover, the assumption that the points $\{1,3\}$ are not poles of
$\zeta_\cL(s)$ ensures that $\zeta_{\cD_\cP}(s)$ has simple poles at
these points. It also has a simple pole at $s=\sigma_4(\cP)$ and at all the
poles off the real line by the third assumption. Thus, the spectral triple has  
simple dimension spectrum and the non-negative part of the dimension 
spectrum is given by \eqref{SigmaplusPSC}. The last property about
the poles off the real line follows from the last property of Definition \ref{defan}
and Theorem 2.17 of \cite{Lapidus}.
\endproof


\begin{prop}\label{SpActPSC}
Let $\cP$ be a packing for which the assumptions of Definition \ref{defan} hold.
Then the expansion of the spectral action for the spectral
triple $ST_{PSC}(\cP)$ is of the form
\begin{equation}\label{PSCspact}
\Tr(f(\cD_\cP/\Lambda))\sim \Lambda^3\, \zeta_{\cL}(3)\, f_3 - \Lambda\, \frac{1}{4}\, \zeta_{\cL}(1) \,f_1
+ \Lambda^{\sigma}\, ( \zeta(\sigma-2,\frac{3}{2}) 
- \frac{1}{4} \zeta(\sigma, \frac{3}{2}) )\,  \cR_\sigma \, f_\sigma + \cS_\cP^{osc}(\Lambda),
\end{equation}
where $\sigma=\sigma_4(\cP)$ the packing constant, 
$\cR_\sigma ={\rm Res}_{s=\sigma} \zeta_{\cL}(s)$ the residue of the zeta function of the
fractal string $\cL=\cL(\cP)$, and $f_\beta=\int_0^\infty v^{\beta-1} f(v) dv$, the momenta
of the test function, and $\cS_\cP^{osc}(\Lambda)$ is an oscillatory term involving the
contributions of the points of the dimension spectrum that are off the real line. 
For $R>0$, let $\cS_\cP^{osc}(\Lambda)_{\leq R}$ be the truncation of the
oscillatory terms that only counts the contribution of the off-real poles with
$|\Im(s)|\leq R$. Then the oscillatory term $\cS_\cP^{osc}(\Lambda)$ 
can be approximated by a sequence
\begin{equation}\label{oscnseries}
 \cS_\cP^{osc}(\Lambda)_{\leq R} \sim  \sum_{j=0}^{N_n} \Lambda^{\sigma_{n,j}} \ff_{\sigma_{n,j}}(\theta_n(\Lambda)), 
\end{equation} 
where $n\to \infty$ as $R\to \infty$, and where $\sigma_{n,j} =\Re(s_{n,j,m})$, for 
$$\{ s_{n,j,m} = \sigma_{n,j} + i (\alpha_{n,j} + \frac{2\pi m}{\log b_n} ) \}_{j=0,\ldots, N_n, m\in \Z}$$ 
the set of non-real poles of the
zeta functions $\zeta_{\cL_n}(s)$, with $\sigma_{min}\leq \sigma_{n,j} \leq \sigma$ and
periods $2\pi/ \log b_n \to \infty$ and $n\to \infty$. The $\ff_{\sigma_{n,j}}(\theta_n(\Lambda))$
are smooth functions of the circle variable $\theta_n= \frac{\log\lambda}{\log b_n}$, with
Fourier expansion $\ff_{\sigma_{n,j}}(\theta_n)= \sum_m f_{s_{n,m}} e^{2\pi i m \theta_n}$ with
$$ f_{s_{n,j,m}} =(\zeta(s_{n,j,m}-2,\frac{3}{2})-\frac{1}{4}\zeta(s_{n,j,m},\frac{3}{2})) \, {\rm Res}_{s=s_{n,j,m}} \zeta_{\cL_n}(s) \,\, \int_0^\infty f(u) \, u^{s_{n,j,m}-1}\, du. $$
\end{prop}

\proof Under the assumptions listed in Definition \ref{defan} on the zeta function $\zeta_{\cL}(s)$,
the residues at the points $s=1$ and $s=3$ of the dimension spectrum are
given, respectively, by 
$$ {\rm Res}_{s=1}\zeta_{\cD_{\cP}}(s)=\frac{-1}{2} {\rm Res}_{s=1} \zeta(s,\frac{3}{2})\cdot \zeta_{\cL}(s) =-\frac{1}{2}\, \zeta_{\cL}(1) $$
$$ {\rm Res}_{s=3}\zeta_{\cD_{\cP}}(s)=2 {\rm Res}_{s=3} \zeta(s-2,\frac{3}{2}) \cdot \zeta_{\cL}(s) = 2\, \zeta_{\cL}(3).  $$
Thus, the terms in the expansion of the spectral action
$$ \Tr(f(\cD_\cP/\Lambda))\sim \sum_{\beta \in \Sigma^+_{{\rm ST}_{PSC}}}
f_\beta \Lambda^\beta \cutint |\cD_\cP|^{-\beta}, $$
with $\cutint |\cD_\cP|^{-\beta}$ the residues, as in \eqref{cutint}, 
are given by
$$ \begin{array}{rl}
\Tr(f(\cD_\cP/\Lambda))\sim & 
 \displaystyle{\Lambda^3\, \zeta_{\cL}(3)\,  \int_0^\infty v^2 f(v)\, dv} \\[3mm]
-& \displaystyle{\Lambda\, \frac{1}{4}\, \zeta_{\cL}(1)\,   \int_0^\infty f(v) \, dv} \\[3mm]
+& \displaystyle{\Lambda^{\sigma}\, ( \zeta(\sigma-2,\frac{3}{2}) 
- \frac{1}{4} \zeta(\sigma, \frac{3}{2}) )\,  
{\rm Res}_{s=\sigma} \zeta_{\cL}(s)} \, 
\int_0^\infty v^{\sigma-1} f(v)\, dv, \end{array} $$
where $\sigma=\sigma_4(\cP)$. The approximate form of the oscillatory term
is derived from the last property of Definition \ref{defan} and from the form of
the oscillatory terms of Proposition  \ref{singlescale}. In the case of a self-similar string 
$\cL_n$ with the lattice property, the poles of $\zeta_{\cL_n}(s)$ off the real line
consist of a finite union of sequences of the form 
$s_{n,j,m} = \sigma_{n,j} + i (\alpha_{n,j} + \frac{2\pi m}{\log b_n})$, with
the same period $\frac{2\pi i m}{\log b_n}$ and with $\alpha_{n,0}=0$ and
$\alpha_{n,j}\neq 0$ for $j>0$, and with 
$\sigma_{n,0}=\sigma_n$ the self-similarity dimension, as shown in Theorem 2.17 of
\cite{Lapidus}. The terms in the expansion of the spectral
action that correspond to these poles are then approximated, for large $n$,
by a finite sum of terms as in Proposition  \ref{singlescale}.
\endproof

For the purpose of this paper we will not give a more detailed analysis of
the convergence of the approximation by the sequence 
$\sum_j \Lambda^{\sigma_{n,j}} \ff_{\sigma_{n,j}}(\theta_n(\Lambda))$.
A more precise analytic discussion of the nature of the approximation in \eqref{oscnseries} 
will require a more detailed understanding of self-similar structures in 
higher-dimensional Apollonian sphere packings than is presently available,
and will need to be addressed elsewhere.
In terms of the expansion of the spectral action we are going to use in explicit 
gravitational models, we will truncate the series of oscillatory terms as discussed in 
\S \ref{ApproxSec}.

\medskip
\subsection{Zeta regularization}\label{zetareg}

The expression \eqref{PSCspact} for the spectral action of the sphere packing should be
regarded as a ``zeta regularized" form of the divergent series
$$ \cS_\cP(\Lambda) = \sum_{k=0}^\infty \cS_{S^3_{a_k}}(\Lambda) $$
that adds the contributions coming from the spectral actions of
the individual spheres in the packing. Indeed, since the spectral action of an
individual sphere is of the form
\begin{equation}\label{Sonesphere}
\cS_{S^3_{a_{n,k}}} (\Lambda) = 
 \Lambda^3\,  a_{n,k}^3\, f_3 -\frac{1}{4} \Lambda\, a_{n,k}\, f_1 + O((\Lambda a_{n,k})^{-K}), 
\end{equation} 
 and both points $1$ and $3$ are smaller than the exponent of convergence $\sigma_4(\cP)$
 of the series $\sum_{n,k} a_{n,k}^s$, the series
 \begin{equation}\label{13divseries}
  \Lambda^3 f_3 \sum_{n,k} a_{n,k}^3 - \frac{1}{4} \Lambda f_1 \sum_{n,k} a_{n,k} 
 \end{equation} 
 is divergent and requires a suitable regularization. The spectral action \eqref{PSCspact}
 can be interpreted as such a regularization. Notice also that the error term $O((\Lambda a_{n,k})^{-K})$
 is very small for a fixed radius $a_{n,k}$ and for sufficiently large $\Lambda$, but when the
 radii $a_{n,k}$ vary over the set $\cL(\cP)$ of lengths of the packing $\cP$ it becomes large for
 any given $\Lambda$, so that \eqref{Sonesphere} cannot be extended directly to the whole packing. 
 The term 
 \begin{equation}\label{zetareg13}
 \Lambda^3 f_3 \, \zeta_{\cL}(3) -\frac{1}{4} \Lambda f_1\, \zeta_{\cL}(1) 
 \end{equation}
 in \eqref{PSCspact} is just a classical form of zeta regularization of the series \eqref{13divseries},
 with the divergent $\sum a_{n,k}^3$ replaced by $\zeta_{\cL}(3)$ and the divergent
 $\sum a_{n,k}$ replaced by $\zeta_{\cL}(1)$. The additional term in \eqref{PSCspact},
 which depends on the residue of $\zeta_{\cL}(s)$ at $s=\sigma_4(\cP)$ detects the presence
 of a fractal structure in the geometry.
 We discuss these issues further in \S \ref{FractalScaleSec} below.

\medskip
\subsection{Packed Swiss Cheese Spacetime spectral action}\label{spacetimeSec}

In the previous subsection we computed the expansion of the spectral action
for an Apollonian packing of $3$-spheres, under some assumptions on the behavior of
the associated zeta function. Here we consider an associated (Euclidean) spacetime 
model. This generalizes to the case of a packing of spheres the simpler case of a single
sphere $S^3$, where the associated spacetime is just $\R \times S^3$, with the Euclidean
time line $\R$ compactified to a circle $S^1$ of size $\beta$.  We generalize
the form of the spectral action of $S^1_\beta \times S^3_a$, by
replacing the $3$-sphere $S^3_a$ with a packing $\cP$ of $3$-spheres $S^3_{a_{n,k}}$,
and using the results in the previous section.

\begin{prop}\label{SAprodS1P}
Let $\cP$ be a packing of $3$-spheres satisfying the three conditions
of Definition \ref{defan}. Consider the product geometry $S^1_\beta \times \cP$
of $\cP$ with a circle of size $\beta$. Then the spectral action has 
expansion with leading terms of the form
\begin{equation}\label{SAPS1}
\begin{array}{ll}
\cS_{S^1_\beta \times \cP}(\Lambda) \sim & \displaystyle{2 \beta \left( 
\Lambda^4\, \zeta_{\cL}(3)\, \fh_3 - \Lambda^2\, \frac{1}{4}\, \zeta_{\cL}(1) \,\fh_1
+ \Lambda^{\sigma+1}\, \left( \zeta(\sigma-2,\frac{3}{2}) 
- \frac{1}{4} \zeta(\sigma, \frac{3}{2}) \right)\,  \cR_\sigma \, \fh_\sigma
\right)} \\[4mm]
& + \cS_{S^1_\beta \times \cP}(\Lambda)^{osc}
\end{array}
\end{equation}
where $\sigma=\sigma_4(\cP)$ is the packing constant \eqref{sigmaD}, $\cR_\sigma$ is the residue
of $\zeta_{\cL}(s)$ at $s=\sigma$, and 
\begin{equation}\label{h1h3}
 \fh_3 :=\pi \int_0^\infty h(\rho^2) \rho^3 d\rho, \ \ \  \fh_1:=2\pi \int_0^\infty h(\rho^2) \rho d\rho. 
\end{equation}
\begin{equation}\label{hsigma}
\fh_\sigma =  2 \int_0^\infty h(\rho^2) \rho^\sigma d\rho.
\end{equation} 
The oscillatory contributions from poles off the real line are approximated by a sequence
\begin{equation}\label{oscnseriesS1}
  \cS_{S^1_\beta \times \cP}(\Lambda)^{osc}_{\leq R} \sim \sum_{j=0}^{N_n} 
  \Lambda^{\sigma_{n,j}+1} \fg_{\sigma_{n,j}}(\theta_n(\Lambda)), 
\end{equation} 
with $n\to \infty$ when $R\to \infty$, 
where $\sigma_{n,j} =\Re(s_{n,j,m})$, with $s_{n,j,m}$ the non-real poles of $\zeta_{\cL_n}(s)$
and $\fg_{\sigma_{n,j}}(\theta_n(\Lambda))$ smooth functions with Fourier coefficients
$$ (\zeta(s_{n,j,m}-2,\frac{3}{2})-\frac{1}{4}\zeta(s_{n,j,m},\frac{3}{2})) {\rm Res}_{s=s_{n,j,m}}\zeta_{\cL_n}(s)
\, \fh_{s_{n,j,m}}, $$
with $\fh_{s_{n,j,m}}$ defined as in \eqref{hsigma}.
\end{prop}
 
\proof 
As observed in Lemma 2 of \cite{CC-uncanny}, the spectral action for 
$S^1_\beta \times S^3_a$, with the Dirac operator
$$ D_{S^1_\beta \times S^3_a} =\left(\begin{matrix} 0 & 
D_{S^3_a} \otimes 1 + i \otimes D_{S^1_\beta}  \\
D_{S^3_a} \otimes 1 - i \otimes D_{S^1_\beta} & 0
\end{matrix}\right) $$
is of the form
\begin{equation}\label{SAS1S3}
 \Tr(h(D^2_{S^1_\beta \times S^3_a}/\Lambda)) \sim 2 \beta \Lambda \Tr(\kappa(D^2_{S^3_a}/\Lambda)), 
\end{equation} 
for a test function $h(x)$, and with the test function $\kappa$ on the right-hand-side satisfying
$\kappa(x^2)=\int_\R h(x^2 + y^2) dy$.
It then follows that the expansion of the 
spectral action on $S^1_\beta \times S^3_a$ is given by (see Theorem 3 of \cite{CC-uncanny})
$$ \Tr(h(D^2_{S^1_\beta \times S^3_a}/\Lambda)) \sim 2\,\beta  \left(\Lambda^4\, a^3\, 
\fh_3 - \frac{1}{4}\,  \Lambda^2\, a\, \fh_1\right), $$
with the notation of \eqref{h1h3}.
We now consider a similar situation, with the product geometry $S^1_\beta \times S^3_a$ replaced 
by $S^1_\beta \times \cP$, where $\cP$ is a packing of $3$-spheres satisfying the conditions
of Definition \ref{defan}. The Dirac operator of the product geometry is again of the form
$$ D_{S^1_\beta \times \cP} =\left(\begin{matrix} 0 & 
\cD_\cP \otimes 1 + i \otimes D_{S^1_\beta}  \\
\cD_\cP \otimes 1 - i \otimes D_{S^1_\beta} & 0
\end{matrix}\right), $$
where $\cD_\cP$ is the Dirac operator of the spectral triple $ST_{PSC}$ described in
Definition \ref{S3defPD}. The same argument as in Lemma 2 of \cite{CC-uncanny}
shows that, as in \eqref{SAS1S3}
$$  \Tr(h(D^2_{S^1_\beta \times \cP}/\Lambda)) \sim 2 \beta \Lambda \Tr(\kappa(\cD^2_\cP/\Lambda)), $$
with the test functions $h$ and $\kappa$ as above. Using the result of Proposition \ref{SpActPSC} we then
obtain, as above, the expression \eqref{SAPS1}, with $\fh_\sigma$ given by
$$ \fh_\sigma =  \int_{\R_+\times \R} x^{\sigma-1} \, h(x^2+y^2)\, dx\,dy 
 =\int_0^\infty h(\rho^2) \rho^\sigma d\rho \int_{-\pi/2}^{\pi/2} \cos(\theta) \, d\theta =
2 \int_0^\infty h(\rho^2) \rho^\sigma d\rho. $$
The structure of the oscillatory terms is obtained as in the previous Proposition.
\endproof

\medskip

In cosmological models based on the spectral action (see \cite{MaPieTeh}, \cite{MaPieTeh2}),
the parameter $\beta$ is an artifact introduced by the choice of a compactification of the
Euclidean time coordinate along a circle of size $\beta$. As discussed in \S 3.1 of
\cite{MaPieTeh2}, the parameter $\beta$ can be interpreted as an inverse temperature and
related to the temperature of the cosmological horizon.

\section{Fractality scale truncation}\label{FractalScaleSec}

A realistic model of fractal structures in cosmology will necessarily involve a choice of scale
at which fractality is cut off: while the universe may involve a fractal structure at the 
scale of galaxy superclusters and clusters, it does not appear fractal at our scales,
hence the self-similarity property is expected to break down at some level. In a gravity
model based on the spectral action, which already naturally involves a dependence
on an energy scale $\Lambda$, it is natural to assume that the scale at which fractality
breaks down will be in some way dependent on $\Lambda$. In the construction of the
spectral triple of the PSCC model, discussed in \S \ref{S3sec} above, we obtained a
spectral action functional as a suitable kind of ``zeta regularization" of the divergent
series  
$$  \sum_{n=0}^\infty \sum_{k=1}^{N_n} \cS_{S^3_{a_{n,k}}}(\Lambda), $$
where the sum is over all the $3$-spheres in the packing $\cP$, with $a_{n,k}$ their
radii, and with $N_n=6\cdot 5^{n-1}$, the number of spheres in the $n$-th
level of the packing construction. Indeed, as we have seen in the previous section, 
the spectral action $\cS_\cP(\Lambda)$ involves a 
zeta regularization of the above series, given by
$$  \left( \Lambda^3 f_3 \sum_{n,k} a_{n,k}^3 - \frac{1}{4} \Lambda f_1 \sum_{n,k} a_{n,k} \right)^{reg} =
 \Lambda^3 f_3 \, \zeta_{\cL}(3) -\frac{1}{4} \Lambda f_1\, \zeta_{\cL}(1) $$
 and an additional term
 $$ \Lambda^{\sigma}\, ( \zeta(\sigma-2,\frac{3}{2}) 
- \frac{1}{4} \zeta(\sigma, \frac{3}{2}) )\,  \cR_\sigma \, f_\sigma $$
involving the residue $\cR_\sigma={\rm Res}_{s=\sigma} \zeta_{\cL}(s)$ at $\sigma=\sigma_4(\cP)$, 
which describes the fractality of the Apollonian packing.

\smallskip
\subsection{Sphere counting function}

In a model where fractality is truncated at a certain scale, one only considers the
sphere packing $\cP$ only up to a certain size. This requires estimating
the number 
\begin{equation}\label{Nalpha}
\cN_\alpha(\cP) = \# \{ S^3_{a_{n,k}}\in \cP \,:\, a_{n,k} \geq \alpha \}
\end{equation}
of spheres in the given packing whose radii are of size at least $\alpha$. In the case
of Apollonian packings of circles and of $2$-spheres it is known, by a result of \cite{Boyd2}, that the
Hausdorff dimension of the residual set of the packing is equal to 
$$ \dim_H (\cR(\cP)) = \lim_{\alpha \to 0} - \frac{\log \cN_\alpha (\cP)}{\log\alpha}, $$
so that, for $\alpha\to 0$, one has $\cN_\alpha(\cP)\sim \alpha^{-\dim_H (\cR(\cP))+o(1)}$,
see also \cite{BPP}. It was proved in \cite{KoOh} that, in fact, one has the stronger result
$\cN_\alpha(\cP)\sim c_\cP \alpha^{-\dim_H(\cR(\cP))}$.
A general heuristic argument for the existence of a power law governing
the behavior of the sphere counting function for sphere packings in arbitrary dimension
is given in \cite{Aste}. Let $\delta(\cP)$ denote the exponent of the power law, so
that, for $\alpha \to 0$ 
\begin{equation}\label{Nalphadelta}
\cN_\alpha(\cP) \sim  \alpha^{-\delta(\cP)+o(1)}.
\end{equation}
In fact, the result of \cite{Boyd2} shows, in the case of an Apollonian packing of circles,
that $\delta(\cP)$ is equal to the packing constant $\sigma_2(\cP)$, which combined with
the result of \cite{Boyd} then gives the identification with the Hausdorff dimension. 
The general argument of \S 2 of \cite{Boyd2} is independent of the dimension an 
it shows that, in general, one has the estimate
\begin{equation}\label{limsupNalpha}
 \limsup_{\alpha\to 0}\,\,  -\frac{\log \cN_\alpha(\cP_D)}{\log \alpha} = \sigma_D(\cP).
\end{equation}
Thus, if the sequence has a limit, then the limit has to be the packing constant $\sigma_D(\cP)$.

\medskip
\subsection{Spectral triple with truncation of fractality scale}

Thus, in a cosmological model where fractality is truncated at a certain size $\alpha$, one
would consider a spectral triple of the form 
$$ (\cA_{\cP_\alpha}, \bigoplus_{n,k\,:\, a_{n,k}\geq \alpha} \cH_{S^3_{a_{n,k}}},  
\bigoplus_{n,k\,:\, a_{n,k}\geq \alpha} D_{S^3_{a_{n,k}}}), $$
where $\cP_\alpha \subset \cP$ is the part of the packing that includes
only those spheres $S^3_{a_{n,k}}$ with $a_{n,k}\geq \alpha$, and
$\cA_{\cP_\alpha}\subset C(\cP_\alpha)$ that satisfies the bounded
commutator condition with the Dirac operator. Correspondingly,
in this case, which involves only finitely many spheres, the spectral action would 
be of the form
\begin{equation}\label{SAPalpha}
\cS_{\cP_\alpha}(\Lambda)= \sum_{n,k\,:\, a_{n,k}\geq \alpha} \cS_{S^3_{a_{n,k}}}(\Lambda).
\end{equation}

\smallskip

\begin{lem}\label{alphadivSA}
Let $\cP$ be a packing of $3$-spheres satisfying the properties of Definition
\ref{defan}, and with the property that the function 
$F(\alpha)=-\log(\cN_\alpha(\cP))/\log(\alpha)$
has a limit for $\alpha\to 0$. Then the spectral action $\cS_{\cP_\alpha}(\Lambda)$ diverges
at least like $\alpha^{-(\sigma_4(\cP)-1)+o(1)}$, when $\alpha \to 0$.
\end{lem}

\proof
Each sphere contributes to the spectral action a term of the form
\begin{equation}\label{SA1ank}
 \Lambda^3 f_3 \sum_{n,k} a_{n,k}^3 - \frac{1}{4} \Lambda f_1 \sum_{n,k} a_{n,k} 
+O((\Lambda a_{n,k})^{-K}). 
\end{equation}
Using the power law \eqref{Nalphadelta} for $\alpha\to 0$ 
we estimate
$$ \sum_{n,k} a_{n,k}^3 \geq \alpha^{3-\delta(\cP)+o(1)} \ \ \ \text{ and } \ \ \
\sum_{n,k} a_{n,k} \geq \alpha^{1-\delta(\cP)+o(1)} . $$
By \eqref{limsupNalpha} and the hypothesis on the existence of the limit, we know 
that $\delta(\cP)=\sigma_4(\cP)$, which we know satisfies $\sigma_4(\cP)>3$, so
that the exponents above are negative. 
Thus, for a fixed $\Lambda$ and for $\alpha\to 0$, the sum
$$ \sum_{n,k\,:\, a_{n,k}\geq \alpha} \left( \Lambda^3 f_3 
\sum_{n,k} a_{n,k}^3 - \frac{1}{4} \Lambda f_1 \sum_{n,k} a_{n,k} \right) $$
diverges at least like the dominant term $\alpha^{-(\delta(\cP)-1)+o(1)}$.
\endproof

\medskip
\subsection{Truncation estimates on the spectral action}\label{truncSec}

One then expects that there will be a good approximation to the spectral action 
$\cS_{\cP_\alpha}(\Lambda)$ of the PSCC obtained
by truncating fractality at a certain $\Lambda$-dependent scale. We discuss
here possible choices of a function $\alpha=\alpha(\Lambda)$ that retain
the property of having a good control on the error term of the spectral
action $\cS_{\cP_{\alpha(\Lambda)}}(\Lambda)$. 
We first recall how one obtains
explicit estimates for the error term in \eqref{SAS3a} of the
spectral action on a $3$-sphere, for a particular class of test
functions, following the argument of 
Corollary 4 of \cite{CC-uncanny}.

\begin{lem}\label{errorterm}
In the case of
a cutoff function of the form  $f(x) = P(\pi x^2)e^{-\pi x^2}$, where $P$ 
is a polynomial of degree $d$, the error term $\epsilon(\Lambda a)$  
in the spectral action
$$ \cS_{S^3_a}(\Lambda)  =
(\Lambda a)^3 \int_\mathbb{R}v^2f(v)dv - \frac{1}{4} (\Lambda a) \int_\mathbb{R}f(v)dv 
+ \epsilon(\Lambda a) $$
satisfies the estimate
\begin{equation}\label{erroreq}
|\epsilon(\Lambda a)| \le (\Lambda a)^3(5+7d+d^2)C_Qe^{-\frac{\pi}{2}(\Lambda a)^2},
\end{equation}
whenever $\Lambda a \ge \sqrt{(d+1)(1+\log (d+1))}$ and $\Lambda a \ge 1$.
The coefficient $C_Q$ to be the sum of the absolute values of the 
coefficients of the polynomial $Q$, with $\hat{f}(x) = Q(\pi x^2)e^{-\pi x^2}$.
\end{lem}

\proof As in Corollary 4 of \cite{CC-uncanny}, we have 
$\hat{f}^{(2)}(x) = (x^2Z_1(\pi x^2) + Z_2(\pi x^2))e^{-\pi x^2}$, 
where $Z_1 = -Q + 2Q' - Q''$, and $Z_2 = \frac{1}{2\pi}(Q - Q')$,
while generally $x^ke^{-x/2} \le 1$, for $x \ge 3k(1+\log k)$. 
For $n \neq 0$ one then has
$$\hat{f}(n\Lambda a) = Q(\pi(n\Lambda a)^2)e^{-\frac{\pi}{2}(n\Lambda a)^2}\cdot 
e^{-\frac{\pi}{2}(n\Lambda a)^2} \le C_Q e^{-\frac{\pi}{2}(n\Lambda a)^2},$$
because, by hypothesis, $\pi (n\Lambda a)^2 \ge 3d(1+\log d)$.
Since the decay is more rapid than simply exponential, we can see that
$$2 \sum_{n=1}^\infty e^{-\frac{\pi}{2}(n\Lambda a)^2}	\le	
2e^{-\frac{\pi}{2}(\Lambda a)^2} + 2\sum_{n=4}^\infty e^{-\frac{\pi}{2}n(\Lambda a)^2}	\le	
2e^{-\frac{\pi}{2}(\Lambda a)^2} + 2\frac{e^{-\frac{3\pi}{2}(\Lambda a)^2}}{-1+e^{\frac{\pi}{2}(\Lambda a)^2}}$$
$$\le 2e^{-\frac{\pi}{2}(\Lambda a)^2} + 2\frac{e^{-\frac{\pi}{2}(\Lambda a)^2} / 23}{-1+4.8}	\le	
2.023e^{-\frac{\pi}{2}(\Lambda a)^2},$$
where we used the assumption that $\Lambda a \ge 1$. Thus, we have
$$\sum_{n \neq 0} \hat{f}(n\Lambda a) \le 2.023C_Q e^{-\frac{\pi}{2}(\Lambda a)^2}. $$
Similarly, for $\hat{f}^{(2)}$ we get
$$\sum_{n \neq 0}\hat{f}^{(2)}(n\Lambda a) \le 2.023(2+3d+\frac{1}{\pi}d^2)
C_Q e^{-\frac{\pi}{2}(\Lambda a)^2}.$$
By looking at the series for the spectral action after applying the Poisson 
summation formula, we see that the above terms contribute to the error as
$$|\epsilon(\Lambda a)| \le (\Lambda a)^3 2.023(2+3d+\frac{1}{\pi}d^2)
C_Q e^{-\frac{\pi}{2}(\Lambda a)^2} + \frac{2.023}{4}\Lambda a C_Q e^{-\frac{\pi}{2}(\Lambda a)^2},$$
which can then be estimated from above as in \eqref{erroreq}.
\endproof

We can then adapt the error estimates of Lemma \ref{errorterm} 
to the PSCC model, by performing a truncation on the 
fractality scale, dependent on the energy scale $\Lambda$.

\smallskip

\begin{rem}\label{alphaLambda}{\rm
Under the assumption that $a_{n,k}\geq \alpha$, the error term in
\eqref{SA1ank} is at most $O((\Lambda \alpha)^{-K})$.
Thus, it is natural to consider a model where the cutoff of fractality 
should happen at a scale $\alpha$ related to $\Lambda$ by
the property that $\alpha(\Lambda) \cdot \Lambda$ grows
like a positive power of $\Lambda$, so that one maintains a
good control on the error term for large $\Lambda$.}
\end{rem}

\begin{prop}\label{alphaLambda2}
Consider a truncated packing $\cP_\alpha$ of $3$-spheres $S^3_{a_{n,k}}$ with
$a_{n,k}\geq \alpha$, where $\alpha=\alpha(\Lambda)=\Lambda^{-1+\gamma}$ for
some $0<\gamma<1$. Then the spectral action, computed using a test
function of the form $f(x)=P(\pi x^2) e^{-\pi x^2}$ with $P$ a polynomial of degree $d$, satisfies
$$ \cS_{\cP_{\Lambda^{-1+\gamma}}}(\Lambda)=  \left(
\sum_{a_{n,k}\geq  \Lambda^{-1+\gamma}}
a_{n,k}^3 \right)  \Lambda^3 f_3 -\frac{1}{4}   \left( \sum_{a_{n,k}\geq  \Lambda^{-1+\gamma}}
a_{n,k} \right) \Lambda f_1 +\epsilon(\Lambda) $$
where the error term satisfies
$$ |\epsilon(\Lambda)|  \leq \Lambda^{3+\sigma(1-\gamma)} a_{\max}^3 (5+7d+d^2) C_Q e^{-\frac{\pi}{2} \Lambda^{2\gamma}}, $$
where $a_{\max}=\max\{ a_{n,k} \}$ is the largest radius in the packing $\cP$, and $\sigma$ is
the packing constant.
\end{prop}

\proof The estimate follows immediately from the previous Lemma, since we
have 
$$ |\epsilon(\Lambda)|  \leq \sum_{a_{n,k}\geq  \Lambda^{-1+\gamma}} \Lambda^3 a_{n,k}^3 (5+7d+d^2)
C_Q e^{-\frac{\pi}{2} (\Lambda a_{n,k})^2} $$
$$ \leq \cN_{\Lambda^{-1+\gamma}}(\cP) \Lambda^3 a_{\max}^3 
 (5+7d+d^2) C_Q e^{-\frac{\pi}{2} \Lambda^{2\gamma}} = \Lambda^{3+\sigma(1-\gamma)} 
 a_{\max}^3 (5+7d+d^2) C_Q e^{-\frac{\pi}{2} \Lambda^{2\gamma}}. $$
 \endproof

\medskip

We consider a further possible way of truncating the spectral action, for
spheres with smaller radii and the behavior of the error term.
We start with the following error term estimate, for a single sphere.

\begin{lem}\label{errortruncate}
For a test function of the form
$f(x) = P(\pi x^2)e^{-\pi x^2}$, for some polynomial $P$ of degree $d$, and for
$M \Lambda a$ the truncation scale, consider the truncated sum
for the spectral action, 
$$ \cS_{S^3_a, M}(\Lambda) = \sum_{\lambda < M \Lambda a} {\rm Mult}(\lambda) \,\,
f(\frac{\lambda}{\Lambda a}) = \sum_{\lambda < M \Lambda a} (\lambda^2 - \frac{1}{4}) 
f(\frac{\lambda}{\Lambda a}). $$
Let $N = \inf \{n \in \frac{1}{2}+\mathbb{Z} \mid n \ge \max\{ M\Lambda a, \frac{3}{2}\} \}$.
Then, assuming that $$\Lambda a \ge \sqrt{(d+1)(1+\log (d+1))} \ \ \ \text{ and } \ \ \ \Lambda a \ge 1, $$ 
the error term satisfies
$$ |\epsilon(\Lambda a)| = 2 \sum_{\lambda \geq N} (\lambda^2-\frac{1}{4}) \,\,
f\left(\frac{\lambda}{\Lambda a}\right) \le \frac{2}{\pi}(\Lambda a)^2(1 
+ \frac{2}{\pi}(\Lambda a)^2)C_P e^{-\frac{\pi}{2}M^2}.
$$
\end{lem}

\proof
Since $u^2-\frac{1}{4}$ is a strictly positive quantity when evaluated at the integers, its absolute 
value is always less than $u^2$. Furthermore, all terms in the series are positive, so that we can 
work with $u^2$ instead of $u^2-\frac{1}{4}$ and our conclusions will remain valid. 
As in Lemma \ref{errorterm}, we use the fact that $x^k e^{-\frac{x}{2}} \le 1$, for all $x \ge 3k(1+\log k)$.
Then, as long as $x \ge \sqrt{(d+1)(1+\log(d+1))}$, we know that
$x^2f(x) \le \frac{1}{\pi}C_P e^{-\frac{\pi}{2}x^2}$.

Let our point of truncation be $M\Lambda a$, so that our sum is over $u < M\Lambda a$. 
Let $N$ be as in the statement. Then we have
$$|\epsilon(\Lambda a)| = 2 \sum_{\lambda \geq N} (\lambda^2-\frac{1}{4})\,\, 
f\left(\frac{\lambda}{\Lambda a}\right) \le 2 \sum_{\lambda\geq N} 
(\Lambda a)^2\left(\frac{\lambda}{\Lambda a}\right)^2 \,
f\left(\frac{\lambda}{\Lambda a}\right)  $$
$$ \le  2 \sum_{\lambda\geq N} (\Lambda a)^2 \frac{1}{\pi}
C_P e^{-\frac{\pi}{2}\left(\frac{\lambda}{\Lambda a}\right)^2}
\leq \frac{2}{\pi}(\Lambda a)^2 C_P 
\sum_{v=0}^\infty e^{-\frac{\pi}{2}\frac{1}{(\Lambda a)^2} (N^2+v)} $$
$$ = \frac{2}{\pi}(\Lambda a)^2 C_P e^{-\frac{\pi}{2}\left(\frac{N}{\Lambda a}\right)^2} \cdot \frac{1}{1-e^{-\frac{\pi}{2}\frac{1}{(\Lambda a)^2}}}
\le \frac{2}{\pi}(\Lambda a)^2C_Pe^{-\frac{\pi}{2}M^2} \cdot \frac{1}{1-e^{-\frac{\pi}{2}\frac{1}{(\Lambda a)^2}}}. $$
Now we check that $(1-e^{-\frac{\pi}{2}\frac{1}{(\Lambda a)^2}})^{-1} < 1 + \frac{2}{\pi}(\Lambda a)^2$. 
Define $g(x) = \frac{1}{1-e^{-1/x}}$. In the case $x \ge 1$, we have
$$e^{-1/x} = \sum_{n=0}^\infty \frac{(-1)^n x^{-n}}{n!} = 1 - x^{-1} + \frac{1}{2}x^{-2} - \sum_{n=2}^\infty \left(\frac{x^{1-2n}}{(2n-1)!} - \frac{x^{-2n}}{(2n)!}\right) $$
$$= 1 - x^{-1} + \frac{1}{2}x^{-2} - 
\sum_{n=2}^\infty \left(x - \frac{1}{2n}\right) \frac{x^{-2n}}{(2n-1)!} \le 1 - x^{-1} + \frac{1}{2}x^{-2} ,$$
since every term in the third sum is positive.
In the case $0 < x < 1$, substituting $u = 1/x$, with $u > 1$, we have 
$$ g(x) = \frac{1}{1-e^{-u}} = \frac{1}{1-e^{-u}} \frac{1+2e^{-u}}{1+2e^{-u}} 
= \frac{1+2e^{-u}}{1 + (e^{-u} - 2e^{-2u})} \le 1 + 2e^{-u}. $$
Compare this with $1+1/u$: the term $2e^{-u}$ decreases faster than 
$1/u$, and $2e^{-1} < 1$, so for all $u > 1$, we have $1+2e^{-u} < 1/u$. Thus, for all $0 < x < 1$
we have $g(x) \le 1 + x$. Together, these two cases give the desired inequality 
$$(1-e^{-\frac{\pi}{2}\frac{1}{(\Lambda a)^2}})^{-1} < 1 + \frac{2}{\pi}(\Lambda a)^2, $$
which then gives the stated estimate for the error term $|\epsilon(\Lambda a)|$. 
\endproof

\smallskip

For a given scale $M$, we consider a summation as above, of the form
$$ \sum_{\frac{3}{2M} < \Lambda a_{n,k} < M} 
\sum_{\lambda = \frac{3}{2}}^{\lceil M\Lambda a_{n,k} \rceil - 1} 2(\lambda^2-\frac{1}{4})
\,\, f\left(\frac{\lambda}{\Lambda a_k}\right). $$
We are interested now in the case of smaller spheres, with $a_{n,k}<\alpha$,
where $\alpha=\alpha(\Lambda)$ was a previously chosen cutoff.  Thus, we relate
the scale $M$ to $\Lambda$ accordingly, by assuming that 
$M=M(\Lambda)=\alpha(\Lambda)\Lambda=\Lambda^\gamma$. This, in turn,
gives the lower bound $\frac{3}{2M} = \frac{3}{2}\Lambda^{-\gamma}$, which
means that we are considering spheres with 
$a_{n,k}\geq \tilde\alpha(\Lambda)= \frac{3}{2}\Lambda^{-(1+\gamma)}$.

\begin{prop}\label{errorM}
For $\alpha(\Lambda)=\Lambda^{-1+\gamma}$ and $\tilde\alpha(\Lambda)=\frac{3}{2}\Lambda^{-(1+\gamma)}$,
and for sufficiently large $\Lambda$, we have 
$$ \cS_{\cP_{\tilde\alpha(\Lambda)}}(\Lambda) =
\sum_{\frac{3}{2}\Lambda^{-\gamma} < \Lambda a_{n,k} < \Lambda^\gamma} 
\sum_{\lambda = \frac{3}{2}}^{\lceil \Lambda^{1+\gamma} a_{n,k} \rceil - 1} 2(\lambda^2-\frac{1}{4})
\,\, f\left(\frac{\lambda}{\Lambda a_k}\right) + \epsilon(\Lambda) , $$
where
$$ |\epsilon(\Lambda)|\leq \left(\frac{3}{2}\right)^\sigma \frac{2}{\pi} 
C_P \Lambda^{2\gamma +(1+\gamma)\sigma} e^{-\frac{\pi}{2} \Lambda^{2\gamma}}. $$
\end{prop}

\proof For large $\Lambda$ we can estimate the number $\cN_{\tilde\alpha(\Lambda)}(\cP)$ 
of spheres with $a_{n,k}\geq \tilde\alpha(\Lambda)$, with 
$\tilde\alpha(\Lambda)^{-\sigma}=\left(\frac{3}{2}\right)^\sigma \Lambda^{(1+\gamma)\sigma}$,
where $\sigma=\sigma_4(\cP)$ is the packing constant. For each sphere in the summation,
we can apply the error term estimate of the previous lemma, with $\Lambda a_{n,k}\leq \Lambda^\gamma$ 
and we obtain
$$ |\epsilon(\Lambda)|\leq \left(\frac{3}{2}\right)^\sigma \Lambda^{(1+\gamma)\sigma} 
\frac{2}{\pi} C_P \Lambda^{2\gamma} (1+\frac{2}{\pi} \Lambda^{2\gamma}) 
 e^{-\frac{\pi}{2} \Lambda^{2\gamma}}. $$
\endproof

We can view the estimate of the error term of this sum as a way to estimate the
effect of changing the cutoff scale from $\alpha(\Lambda)=\Lambda^{-1+\gamma}$
to $\tilde\alpha(\Lambda)=\frac{3}{2}\Lambda^{-(1+\gamma)}$.

\section{Related fractal models}\label{DodeSec}

\subsection{Fractal dodecahedra and cosmic topology}\label{dodecaSec}

In \cite{CMT}, \cite{MaPieTeh}, \cite{MaPieTeh2}, cosmological models based on
the spectral action functional of gravity are constructed for (compactified, Euclidean)
spacetimes of the form $S^1 \times Y$ where $Y$ is either a spherical space form
or a flat Bieberbach manifold, and it is shown that the spectral action detects
the different cosmic topologies through the shape of an associated slow-roll
inflation potential.  In particular, it is shown in \cite{CMT}, \cite{MaPieTeh}, \cite{Teh} 
that the spectral action for a spherical space form $Y=S^3/\Gamma$ is given by
$$ \cS_Y(\Lambda) = \frac{1}{\# \Gamma} \,\cS_{S^3}(\Lambda), $$
independently of the spin structure (even though the Dirac spectrum itself changes
for different spin structures). Of particular interest for cosmic topology is the
case where $Y$ is the Poincar\'e homology $3$-sphere (dodecahedral space),
which is regarded as one of the most promising candidates for a non-simply
connected cosmic topology, \cite{Caill}, \cite{Luminet}.

\smallskip

We consider here a different possible model with fractal structure, where the
building blocks are spherical dodecahedra, folded up to form Poincar\'e homology
spheres, arranged in a fractal configuration that generalizes the Sierpi\'nski fractal to
dodecahedral geometry. Other similar constructions can be done using other solids,
and adapted to the other spherical space form candidates for cosmic topologies.
These fractals are much simpler in structure than the Apollonian sphere
packing described in the previous section, as the successive levels of the
construction are all obtained by repeatedly applying the same uniform contraction
factor. This makes the computation of the Hausdorff dimension immediate, as well
as its identification with the exponent of convergence of the relevant zeta function.
Moreover, it is also immediately clear that an analytic continuation exists to a
meromorphic function on the entire complex plane, hence all the properties
can be checked more easily.

\smallskip

More precisely, let $Y_a= S^3_a/\cI_{120}$ be the quotient of a round $3$-sphere of
radius $a$ by the isometric action of the icosahedral group $\cI_{120}$.  The choice of
a fundamental domain, given by a spherical dodecahedron, and the action
of the group $\cI_{120}$ determine a tiling of $S^3_a$ consisting of $120$ dodecahedra.
The quotient $3$-manifold $Y_a$ is a Poincar\'e homology sphere, of volume
$Vol(Y_a)=Vol(S^3_a/\cI_{120})=\frac{\pi^2}{60} a^3$.

\smallskip

Consider now the following well known construction of a Sierpi\'nski type fractal based
on the dodecahedron. Starting with an initial (solid) regular dodecahedron, one replaces it with
$20$ new regular dodecahedra, contained inside the volume bounded by the initial one,
each placed in the corner corresponding to one of the $20$ vertices of the original
dodecahedron. Each of the new dodecahedra is a copy of the original one scaled
by a factor of $(2+\phi)^{-1}$ where $\phi$ is the golden ratio. One keeps iterating
this procedure on each of the dodecahedra in the new configuration. Let $\cP_{Y,n}$ 
be the union of dodecahedra obtained at the $n$-th step of the construction, where
we simultaneously perform the identification of faces, in each dodecahedron, according
to the action of $\cI_{120}$, so that each is folded up into a Poincar\'e homology $3$-sphere.
Let $\cP_Y$ denote the resulting limit, which in the set theoretic sense is given by the
intersection $\cP_Y=\cap_{n\geq 1} \cP_{Y,n}$. The following fact then follows immediately.

\begin{lem}
The Hausdorff dimension of the resulting set is 
\begin{equation}\label{dimHdodeca}
\dim_H \cP_Y=\frac{\log(20)}{\log(2+\phi)} = 2.3296...
\end{equation}
This is equal to the exponent of convergence of the series 
$$ \zeta_{\cL(\cP_Y)}(s)=\sum_{n\geq 0} 20^n (2+\phi)^{-ns}, $$
which has analytic continuation to a meromorphic function on $\C$,
$$ \zeta_{\cL(\cP_Y)}(s)= \frac{1}{1-20 (2+\phi)^{-s}} $$
with simple poles at the points 
\begin{equation}\label{polesdodeca}
s_m=\frac{\log(20)}{\log(2+\phi)}+ \frac{2\pi i m}{\log(2+\phi)}, \ \ \text{ with } \ \  m\in \Z,
\end{equation}
all with the same residue
\begin{equation}\label{icoRessm}
{\rm Res}_{s=s_m} \zeta_{\cL(\cP_Y)}(s) = \frac{1}{\log(2+\phi)}.
\end{equation}
\end{lem}

We then construct a spectral triple for the configuration $\cP_Y$ using the
same procedure as in \cite{CIL}. Let $(C(Y_a),\cH_{Y_a}, D_{Y_a})$ be 
the spectral triple of $Y_a$ with $\cH_{Y_a}=L^2(Y_a,\bS)$ the square integrable
spinors and $D_{Y_a}$ the Dirac operator. See \cite{MaPieTeh}, \cite{Teh} for a 
more detailed discussion of these data.

\begin{prop}\label{S3dodeca}
The spectral triple  
$$ (\cA_{\cP_Y},\cH_{\cP_Y}, \cD_{\cP_Y}) = (\cA_{\cP_Y},\oplus_n \cH_{Y_{a_n}}, \oplus_n D_{Y_{a_n}}), $$
with $\cA_{\cP_Y}\subset C(\cP_Y)$
satisfying the bounded commutator condition, and with
$a_n=a (2+\phi)^{-n}$, has zeta function 
\begin{equation}\label{zetaDdodeca}
\zeta_{\cD_{\cP_Y}}(s) = \frac{a^s}{120} \left( 2 \zeta(s-2,\frac{3}{2})-\frac{1}{2}\zeta(s,\frac{3}{2}) \right)
\zeta_{\cL(\cP_Y)}(s).
\end{equation}
The positive part of the dimension spectrum is $\Sigma^+=\{ 1, \sigma, 3 \}$, with
$\sigma = \dim_H(\cP_Y)$, while the full dimension spectrum $\Sigma$ also contains
the points \eqref{polesdodeca} off the real line. The metric dimension of the spectral triple
is $3$.
\end{prop}

\proof The spectrum of the Dirac operator on the Poincar\'e homology sphere $Y_a$, with
the correct multiplicities, can be computed explicitly using the method of \cite{Baer} of
generating functions, see \cite{Teh}. In the case of the trivial spin structure, it is shown in
\cite{Teh} that one can separate the spectrum into $60$ arithmetic progressions $\{ \frac{3}{2} +k + 60j \}$
with multiplicities interpolated by $60$ explicit polynomials $P_k(\frac{3}{2} +k + 60j)={\rm Mult}(\frac{3}{2} +k + 60j)$, 
which satisfy 
$$ \sum_{k=0}^{59} P_k(u)= \frac{1}{2} u^2 - \frac{1}{8}. $$
This implies that 
$$ \cS_{Y_a}(\Lambda)= \sum_{k=0}^{59} \sum_{j\in \Z} P_k(\frac{3}{2} +k + 60j) f(\frac{\frac{3}{2} +k + 60j}{\Lambda})
=  \frac{1}{120} \cS_{S^3_a}(\Lambda)  $$
and that
$$  \zeta_{D_{Y_a}}(s)= \sum_{k=0}^{59} \sum_{j\in \Z} P_k(\frac{3}{2} +k + 60j)\,  \left| \frac{3}{2} +k + 60j \right|^{-s} =
\frac{1}{120}\, \zeta_{D_{S^3_a}}(s), $$
hence \eqref{zetaDdodeca} then follows as in Proposition \ref{zetaST}.
\endproof

\smallskip

Using the same technique that we used in the previous construction, we can
then compute the leading terms in the expansion of the spectral action. 
In this case we have a completely explicit description of the poles off the
real line, so we also obtain a more explicit description of the log oscillatory
corrections to the spectral action, which in this model behaves exactly as in 
the case of Proposition \ref{singlescale}. 
\begin{prop}\label{SAdodeca}
Let $\cP_Y$ be the fractal arrangement described above, to which we
assign a spectral triple as in Proposition \ref{S3dodeca}. 
The contribution to the expansion of the spectral action $\Tr(f(\cD_{\cP_Y}/\Lambda))$ 
coming from the points of the dimension spectrum that lie on the positive real line is given by
\begin{equation}\label{SAdod}
   (\Lambda a)^3 \frac{\zeta_{\cL(\cP_Y)}(3)}{120} f_3 
 - \Lambda a \frac{\zeta_{\cL(\cP_Y)}(1)}{120} f_1 + (\Lambda a)^\sigma
 \frac{ \zeta(\sigma-2,\frac{3}{2})-\frac{1}{4} \zeta(\sigma,\frac{3}{2}) }{120 \log(2+\phi)} f_\sigma,
\end{equation}
where $\sigma =\dim_H(\cP_Y)$, while the contribution to the expansion of
the spectral action coming from the non-real points of the dimension spectrum
is given by a Fourier series
$$ \Lambda^\sigma \sum_{m\neq 0} \Lambda^{\frac{2\pi i m}{\log(2+\phi)} }
\frac{ \zeta(s_m-2,\frac{3}{2})-\frac{1}{4} \zeta(s_m,\frac{3}{2}) }{120 \log(2+\phi)} f_{s_m}, $$
with
$$ f_{s_m} = \int_0^\infty f(u) u^{s_m-1}\, du. $$
The series converges absolutely to a smooth function of the angular variable 
$\theta=\frac{\log\Lambda}{\log(2+\phi)}$ mod $2\pi\Z$.
\end{prop}

\proof The argument is exactly as in Proposition \ref{SpActPSC}, where we
now have $2<\sigma<3$ and the residue $\cR_\sigma$ (as well as those at the poles
off the real line) is given by
$$ \cR_\sigma ={\rm Res}_{s=\sigma} \zeta_{\cL(\cP_Y)}(s) = \frac{1}{\log(2+\phi)}.  $$
The complex poles $s_m$ of $\zeta_{\cL}(s)$ lie within the region of absolute
convergence of the series defining the Hurwitz zeta function, hence the size of the terms
$|\zeta(s_m-2,\frac{3}{2})-\frac{1}{4} \zeta(s_m,\frac{3}{2})|$ is controlled by a term
$|\zeta(\sigma-2,\frac{3}{2})|+\frac{1}{4} |\zeta(\sigma,\frac{3}{2})|$. The convergence
of the Fourier series above is then controlled by the convergence of $\sum_m 
\Lambda^{\frac{2\pi i m}{\log(2+\phi)}} f_{s_m}$, which
converges absolutely to a smooth function of the periodic angle variable 
$\theta$, as shown in Proposition \ref{singlescale}.
\endproof

\smallskip

\begin{cor}\label{PY4dim}
Consider a (Euclidean, compactified) spacetime model of the form 
$S^1_\beta \times \cP_Y$, with $\cP_Y$ the fractal arrangement as
above, and with $\beta$ the size of the compactification. Then the
contribution to the expansion of the spectral action $\cS_{S^1_\beta \times \cP_Y}(\Lambda)$ coming
from real points of the dimension spectrum is given by
\begin{equation}\label{SAS1dod}
  2\beta \left( \Lambda^4  \frac{a^3 \zeta_{\cL(\cP_Y)}(3)}{120} \fh_3 
 - \Lambda^2 \frac{a \zeta_{\cL(\cP_Y)}(1)}{120} \fh_1 + \Lambda^{\sigma+1} 
 \frac{ a^\sigma (\zeta(\sigma-2,\frac{3}{2})-\frac{1}{4} \zeta(\sigma,\frac{3}{2})) }{120 \log(2+\phi)} 
 \fh_\sigma \right),
\end{equation}
with $\sigma =\dim_H(\cP_Y)$ and $\fh_1,\fh_3,\fh_\sigma$ as in \eqref{h1h3} and
\eqref{hsigma}. The contribution to the points of the dimension spectrum that are
off the real line is a Fourier series
$$ \Lambda^{\sigma+1} \frac{2\beta a^\sigma}{120 \log(2+\phi)} \sum_{m\neq 0} (\Lambda a)^{\frac{2\pi i m}{\log(2+\phi)}} (\zeta(s_m-2,\frac{3}{2})-\frac{1}{4} \zeta(s_m,\frac{3}{2})) \fh_{s_m}, $$
where the coefficients $\fh_{s_m}$ are given by
$$ \fh_{s_m} = 2\int_0^\infty h(\rho^2) \rho^{s_m} d\rho. $$
The series converges to a smooth function of $\theta=\frac{\log\Lambda}{\log(2+\phi)}$ mod $2\pi$.
\end{cor}

\proof The result follows exactly as in Proposition \ref{SAprodS1P}.
\endproof

\medskip
\section{Slow roll inflation in fractal universes}\label{InflSec}

In the case of a compactified Euclidean spacetime of the form
$S^1_\beta \times S^3_a$, it was shown in \cite{CC-uncanny}
that perturbations of the Dirac operator by a scalar field
$D^2\mapsto D^2+\phi^2$ produce, in the calculation
of the spectral action, a potential $V(\phi)$ for the scalar field, obtained
as a combination of functions $\cV(\phi^2/\Lambda^2)$ and $\cW(\phi^2/\Lambda^2)$
with
$$ \Tr(h((D^2 + \phi^2)/\Lambda^2))) \sim \pi \Lambda^4 \beta a^3 \int_0^\infty u h(u)du
- \frac{\pi}{2} \Lambda^2 \beta a \int_0^\infty h(u) du $$ $$
+ \pi \Lambda^4 \beta a^3\, \cV(\phi^2/\Lambda^2) +\frac{1}{2} \Lambda^2 \beta a\, \cW(\phi^2/\Lambda^2), 
$$
where the functions are of the form
$$ \cV(x)= \int_0^\infty u (h(u+x)-h(u))du, \ \ \ \ \ \cW(x)= \int_0^x h(u)du . $$

\smallskip

In \cite{CC-uncanny} it was first observed that the potential obtained in this
way has the typical shape of the slow-roll inflation potentials. Whether one
can accommodate a satisfactory inflaton model within the spectral action
paradigm is still debated. It was first proposed that the Higgs field might
play the role of inflaton field, but the possibility of a Higgs based inflation
scenario in the noncommutative geometry model was ruled out as 
incompatible with the measured value of the top quark mass in \cite{BFS}, 
based on constraints coming from the CMB data. In the noncommutative
geometry models of gravity coupled to matter, the Higgs sector arises as 
inner fluctuation of the Dirac operator in the non-commutative fiber directions
of an almost-commutative geometry. By contrast, even in the pure gravity case,
where there is no finite non-commutative geometry, one can introduce a
scalar perturbation of the Dirac operator of the kind described above, 
see the discussion in \S 1.4 of \cite{CMT}.

\smallskip

It was observed in \cite{MaPieTeh}, \cite{MaPieTeh2} and \cite{CMT}, where the
construction is generalized for spherical space forms and Bieberbach manifolds,
that the shape of the resulting  slow-roll inflation potential $V(\phi)$ 
distinguishes between (almost all of) the different 
possible topologies and determines detectable signatures of the cosmic topology in
the slow-roll parameters (which in turn determine spectral index and tensor-to-scalar ratio) 
and in the form  of the power spectra for the scalar and tensor fluctuations.
The result is similar in the case of the spherical space forms $Y_a=S^3_a/\Gamma$,
with $V(\phi)$ replaced by $(\# \Gamma)^{-1} V(\phi)$, see \cite{CMT}, \cite{MaPieTeh}.
In this section we discuss how the inflation potential changes in the case of
$S^1_\beta \times \cP$, where $\cP$ is either an Apollonian packing or $3$-spheres
or a configuration based on the fractal dodecahedron and the Poincar\'e homology sphere.
In particular, we show that, in models of inflation based on the spectral action functional, 
the shape of the inflation potential changes depending on the fractal structure, hence
the potential detect measurable effects of the presence and the type of
fractal structure.  In particular, it follows that, in a model of gravity based on the
spectral action, the presence of fractality in the spacetime structure leaves a detectable
signature in quantities, like the slow-roll parameters and the power spectra for the scalar and 
tensor fluctuations, that are in principle measurable in the CMB, modulo the problem of determining
the unknown parameter $\beta$ of the model, already discussed in \cite{MaPieTeh}, 
\cite{MaPieTeh2}. 

\bigskip

We perturb the Dirac operator $D$ of a packing $\cP$ of $3$-spheres by a scalar field $\phi$.
Correspondingly the spectral action is modifies by terms that determine a potential for the
scalar field. We discuss the effect of the real points of the dimension spectrum 
(Proposition \ref{prodS1PV}) and of the fluctuations coming from the points off
the real line (Proposition \ref{prodS1PVflu}) on the shape of the potential.

\begin{prop}\label{prodS1PV}
Let $\cP$ be either a packing of $3$-spheres satisfying the three conditions
of Definition \ref{defan}, or a configuration of Poincar\'e homology $3$-spheres arranged
according to the fractal dodecahedron construction of \S \ref{dodecaSec}. 
Consider the product geometry $S^1_\beta \times \cP$
of $\cP$ with a circle of size $\beta$. 
Then the spectral action satisfies
\begin{equation}\label{SAVPS1}
\begin{array}{ll}
 \Tr(h((D^2 + \phi^2)/\Lambda^2))) \sim &  2 \beta \left( 
\Lambda^4\, \zeta_{\cL}(3)\, \fh_3 - \Lambda^2\, \frac{1}{4}\, \zeta_{\cL}(1) \,\fh_1 \right) \\[3mm]
& + 2 \beta \Lambda^{\sigma+1}\, \left( \zeta(\sigma-2,\frac{3}{2}) 
- \frac{1}{4} \zeta(\sigma, \frac{3}{2}) \right)\,  \cR_\sigma \, \fh_\sigma
\\[3mm]
& + \pi \Lambda^4 \beta \zeta_\cL(3)\, \cV(\phi^2/\Lambda^2) +\frac{\pi}{2} \Lambda^2 \beta \zeta_\cL(1)\, \cW(\phi^2/\Lambda^2) \\[3mm]
& + 4 \beta \Lambda^{\sigma +1} \left( \zeta(\sigma-2,\frac{3}{2}) 
- \frac{1}{4} \zeta(\sigma, \frac{3}{2}) \right)\,  \cR_\sigma \, \cU_\sigma (\phi^2/\Lambda^2) \\[3mm]
& + \cS_{D,\phi}^{osc}(\Lambda),
\end{array}
\end{equation}
where the last term collects the fluctuations coming from the log-oscillatory terms contributed
by the poles of the zeta functions that are off the real line, described in Proposition \ref{prodS1PVflu} below.
The potentials $\cV$, $\cW$, and $\cU_\sigma$ are, respectively, given by
\begin{equation}\label{VWpot}
\cV(x)= \int_0^\infty u (h(u+x)-h(u))du, \ \ \  \ \  \cW(x)= \int_0^x h(u)du, 
\end{equation}
\begin{equation}\label{Upot}
 \cU_\sigma (x)=\int_0^\infty u^{(\sigma-1)/2}  (h(u+x)-h(u))du,
\end{equation}
where $\sigma=\sigma_4(\cP)$ is the packing constant \eqref{sigmaD}, $\cR_\sigma$ is the residue
of $\zeta_{\cL}(s)$ at $s=\sigma$, and $\fh_1$, $\fh_3$, $\fh_\sigma$ are as in \eqref{h1h3}, \eqref{hsigma}.
 \end{prop}
 
 \proof The argument follows directly from Proposition \ref{SAprodS1P}, along the lines of 
 Theorem 7 of \cite{CC-uncanny}. We have
 $$ \int_0^\infty h(\rho^2) \rho^3 d\rho = \frac{1}{2} \int_0^\infty u h(u) du, $$
 which gives rise to the term
 $$ \int_0^\infty u (h(u+x)-h(u)) du $$
 in the $\cV$ part of the potential; similarly, we have
 $$ \int_0^\infty h(\rho^2) \rho d\rho = \frac{1}{2} \int_0^\infty h(u) du, $$
 which gives the term
 $$ \int_0^x h(u) du $$
 in the $\cW$ part of the potential, and the term
 $$ \int_0^\infty h(\rho^2) \rho^\sigma d\rho = \frac{1}{2} \int_0^\infty u^{\frac{\sigma-1}{2}} h(u) du, $$
 which gives the term
 $$ \int_0^\infty u^{\frac{\sigma-1}{2}} (h(u+x)-h(u)) du $$
 in the $\cU_\sigma$ part of the potential.
 \endproof
 
 \smallskip
 
 \begin{prop}\label{prodS1PVflu}
 Under the assumptions of Definition \ref{defan}, the fluctuation terms $\cS_{D,\phi}^{osc}(\Lambda)$ 
 are of the form $\cS_{D,\phi}^{osc}(\Lambda) = \cS_{S^1_\beta\times \cP}(\Lambda)^{osc} +
 \bU_\sigma^{osc}(\phi)$, where $\cS_{S^1_\beta\times \cP}(\Lambda)^{osc}$ is as in 
 \eqref{oscnseriesS1}. Let $\bU_\sigma^{osc}(\phi)_{\leq R}$ be the approximation to 
 $\bU_\sigma^{osc}(\phi)$ that only counts the contribution of non-real poles with $|\Im(s)|\leq R$.
 The oscillatory term $\bU_\sigma^{osc}(\phi)_{\leq R}$ is approximated by a sequence
 \begin{equation}\label{Uoscphi}
 4\beta \sum_{j=1}^{N_n} \Lambda^{\sigma_{n,j}+1} \sum_m \Lambda^{i (\alpha_{n,j}+\frac{2\pi  m}{\log b_n})} (\zeta(s_{n,j,m}-2,\frac{3}{2})-\frac{1}{4} \zeta(s_{n,j,m},\frac{3}{2})) \Res_{s=s_{n,j,m}}\zeta_{\cL_n}(s) \, \cU_{s_{n,j,m}}(\phi^2/\Lambda^2),
 \end{equation}
 where $n\to \infty$ and $R\to \infty$, with
 $$ \cU_{s_{n,j,m}}(x)=\int_0^\infty u^{(s_{n,j,m}-1)/2}  (h(u+x)-h(u))du, $$
 and where $s_{n,j,m}=\sigma_{n,j} + i (\alpha_{n,j} + \frac{2\pi  m}{\log b_n})$ are the complex zeros
 of the series of self-similar strings $\cL_n$ with the lattice property approximating the complex
 poles of $\zeta_{\cL(\cP)}(s)$.
 \end{prop}
 
 \proof The result is obtained as in Proposition \ref{prodS1PV}, using the results 
 of Proposition \ref{SAprodS1P}. 
 \endproof
 
 \smallskip
 
The slow-roll inflation potential $V(\phi)$ is obtained from the combination of functions $\cV$, $\cW$, $\cU$
that appears in the expansion of the spectral action above.

\begin{cor}\label{inflV}
The function 
$$ \pi \Lambda^4 \beta \zeta_\cL(3)\, \cV(\phi^2/\Lambda^2) +\frac{\pi}{2} \Lambda^2 \beta \zeta_\cL(1)\, \cW(\phi^2/\Lambda^2) $$ $$ +4 \beta \Lambda^{\sigma +1} \left( \zeta(\sigma-2,\frac{3}{2}) 
- \frac{1}{4} \zeta(\sigma, \frac{3}{2}) \right)\,  \cR_\sigma \, \cU_\sigma (\phi^2/\Lambda^2) 
+\bU_\sigma^{osc}(\phi) $$
depends explicitly on the presence of fractality, through the coefficients $\zeta_\cL(3)$, $\zeta_\cL(1)$,
the residue $\cR_\sigma$, the packing constant $\sigma$, and the oscillatory fluctuations.
\end{cor}

\smallskip

\begin{cor}\label{inflVdodeca}
In the case of the fractal arrangement $\cP_Y$ of dodecahedral spaces considered
in the previous section, the form of the fluctuations in the inflation potential is simpler
and given by the Fourier series
$$ \bU_\sigma^{osc}(\phi)= \frac{4\beta \Lambda^{\sigma+1}}{\log(2+\phi)}
\sum_m \left(\zeta(\sigma+\frac{2\pi i m}{\log(2+\phi)}-2, \frac{3}{2}) - \frac{1}{4} \zeta(\sigma+\frac{2\pi i m}{\log(2+\phi)}, \frac{3}{2})\right) \cU_{s_m}(\frac{\phi^2}{\Lambda^2}), $$
with
$$ \cU_{s_m}(x)= \int_0^\infty u^{(\sigma+\frac{2\pi i m}{\log(2+\phi)}-1)/2} (h(u+x)-h(u))\, du. $$
\end{cor}

\proof This follows directly from the results of the previous section, by proceeding as
in the proof of Proposition \ref{prodS1PV}.
\endproof

\section{Conclusions and further questions}

\subsection{Conclusions}
In this paper we considered a model of gravity based on the spectral action functional.
This is known to recover, via its asymptotic expansion, the usual Einstein-Hilbert action
with cosmological term, along with modified gravity terms (conformal and Gauss--Bonnet gravity).
We considered simple models of (Euclidean, compactified) spacetimes of the form
$S^1_\beta \times \cP$, where $\beta$ is the size of the $S^1$-compactification and
$\cP$ is a fractal configuration built out of $3$-spheres (Apollonian packings) or
of other spherical space forms (Sierpi\'nski fractals). We evaluated the leading terms
of the expansion of the spectral action, using information on the zeta function
of the Dirac operator of a spectral triple, and we compared them, respectively, with 
the corresponding terms in the simpler case of $S^1_\beta \times S^3_a$ (spatial
sections given by a single sphere of radius $a$) or $S^1_\beta \times Y$ where
$Y$ is a spherical space form, in particular the Poincar\'e homology sphere (dodecahedral space).
We regard the case of $S^1_\beta \times \cP$, where $\cP$ is an Apollonian
packing of $3$-spheres or a configuration obtained from a Sierpi\'nski fractal dodecahedron, 
as a model of possible presence of fractality in spacetime geometry: a version of
Packed Swiss Cheese Cosmology models. We showed that the resulting leading terms of the
expansion of the spectral action for $S^1_\beta \times \cP$ differ from 
those of the ordinary $S^1_\beta \times S^3$ (or $S^1_\beta \times Y$)
case in the following ways:
\begin{itemize}
\item The term $2 \Lambda^4 \beta a^3 \fh_3 - \frac{1}{2} \Lambda^2 \beta a \fh_1$,
respectively corresponding to the cosmological and the Einstein--Hilbert term, are
replaced by terms of the form $2 \Lambda^4  \beta \zeta_{\cL}(3) \fh_3 
- \frac{1}{2} \Lambda^2 \beta \zeta_{\cL}(1) \fh_1$, which can be seen as a zeta
regularization of the divergent series of the $3$-sphere terms summed over the packing.
\item There is an additional term in the gravity action functional of the form 
$$ \Lambda^{\sigma+1} \left( \zeta(\sigma-2,\frac{3}{2})-\frac{1}{4} \zeta(\sigma,\frac{3}{2})\right) \cR_\sigma \fh_\sigma, $$
where $\zeta(s,x)$ is the Hurwitz zeta function and $\sigma$ is the packing constant of $\cP$, 
with $\cR_\sigma$ the residue at $\sigma$ of
the zeta function $\zeta_{\cL}(s)$ of the fractal-packing. For a fractal dodecahedron $\sigma$
is the Hausdorff dimension, while for an Apollonian packing it is conjecturally
the Hausdorff dimension of the residual set. In both cases this term detects modifications to the
gravity action functional due to the presence of a fractal structure.
\item This additional term is further corrected by a Fourier series of log-oscillatory fluctuations,
coming from the presence of points of the dimension spectrum off the real line (another
purely fractal phenomenon).
\item The perturbation $D^2 \mapsto D^2 +\phi^2$ of the Dirac operator determines a slow-roll
inflation potential $V(\phi)$ for the field $\phi$. The shape of the potential 
detects the presence of fractality through the coefficients $\zeta_\cL(3)$, $\zeta_\cL(1)$,
the packing constant $\sigma$, and the residue at $\sigma$ of $\zeta_\cL(s)$, and oscillatory
fluctuations.
\end{itemize}

\smallskip
\subsection{Further questions}

There are a number of questions, both mathematical and physical, that arise in
relation to improving the model of gravity, based on the spectral action, on
cosmologies exhibiting fractality. On the mathematical side, as we have seen, one needs
a better understanding of the properties of higher-dimensional Apollonian packings,
especially with respect to characterizing the presence of exact self-similarity, extending
results like \cite{ChiDo} beyond the $2$-dimensional case of circle packings. 

\smallskip

{}From the physics viewpoint, the logic we followed in this paper is along
the lines of several other recent results, where one considers a classical model
of (Euclidean, compactified) spacetime and computes what the expansion of
the spectral action looks like, either with the full infinite series of the asymptotic
expansion, or at least with the leading terms up to order $\Lambda^0$, see for
instance \cite{CC-uncanny}, \cite{CC-rw}, \cite{FFM}, \cite{FGK}, \cite{Teh}. Under
this perspective, one can consider the classical Packed Swiss Cheese Cosmologies.
These originate in a spacetime model introduced in the late '60s in \cite{Rees}
as a prototype model of isotropic but non-homogeneous spacetimes (as opposed
to, for instance, the Bianchi IX examples of homogeneous non-isotropic spaces).
In other words, the original construction of the Packed Swiss Cheese Cosmology
is dictated by imposing certain kind of regularity requirements on the geometry.
These classical models have then been tested as possible models of fractal
and multifractal structures in spacetime. Here we investigate how the spectral
action functional, seen as our choice of action functional of gravity, behaves on
this specific classical geometry. A first important question in this direction, which
would make the model more realistic (closer to the original construction of
\cite{Rees}) would be to start from a (Euclidean, compactified) Robertson--Walker 
spacetime and carve out balls, so as to obtain a residual Apollonian sphere
packing, rather than adopting the simplified model we considered here of a 
product $\cP \times S^1$. In terms of computations of the spectral action, this
would mean adapting the computation for the Roberston--Walker metric of
\cite{CC-rw} to the resulting fractal packing, rather than (as we did here)
adapting the computations of \cite{CC-uncanny} for $S^3$ to the
case of the Apollonian packings of $3$-spheres, or the fractal dodecahedral packing of
Poincar\'e homology spheres. A second question would be to adopt a different
viewpoint and derive spacetime models (possibly with some form of
fractal structures or of noncommutativity) from a {\em least action principle}
applied to the spectral action functional. More explicitly, the question would
be whether such a variational principle can be reinterpreted in terms of
classical theory as a modified gravity model, with an effective stress-energy
tensor (as for instance in the case of $f(R)$-modified gravity). 

\bigskip
\noindent {\bf Acknowledgment} We thank the referees for many extremely useful comments
and suggestions that greatly improved the paper. The first author was supported 
by a Summer Undergraduate Research Fellowship at Caltech. The second
author is supported by NSF grants DMS-1007207, DMS-1201512, PHY-1205440, and
by the Perimeter Institute for Theoretical Physics.


\begin{thebibliography}{99}

\bibitem{Aste} T.~Aste, {\em Circle, sphere, and drop packing}, Phys. Rev. E, Vol.53 (1996)
N.3, 2571--2579.

\bibitem{Baer}
C.~B\"ar, {\em The Dirac operator on space forms of positive curvature}, 
J. Math. Soc. Japan 48 (1996) N.1, 69--83.

\bibitem{BPP} M.~Borkovec, W.~de Paris, R.~Peikert, {\em The fractal dimension 
of the Apollonian sphere packing}, Fractals 2 (1994) N.4, 521--526.

\bibitem{Boyd} D.~Boyd, {\em The residual set dimension of the Apollonian packing},
Mathematika 20 (1973) 170--174.

\bibitem{Boyd2} D.~Boyd, {\em The sequence of radii of the Apollonian packing}, 
Math. Comp. 39 (1982) 249--254.

\bibitem{BFS} M.~Buck, M.~Fairbairn, M.~Sakellariadou, {\em Inflation in models 
with conformally coupled scalar fields: An application to the noncommutative 
spectral action}, Phys. Rev. D 82 (2010) 043509.

\bibitem{CMT} B.~\'Ca\'ci\'c, M.~Marcolli, K.~Teh, 
{\em Coupling of gravity to matter, spectral action and cosmic topology}, J.
Noncommutative Geometry, Vol.8 (2014) N.2, 473--504.

\bibitem{CC} A.~Chamseddine, A.~Connes, {\em
The spectral action principle}, Comm. Math. Phys. 186 (1997), no. 3, 731--750.

\bibitem{CC-uncanny} A.~Chamseddine, A.~Connes, 
{\em The uncanny precision of the spectral action}, 
Comm. Math. Phys. 293 (2010), no. 3, 867--897. 

\bibitem{CC-rw} A.~Chamseddine, A.~Connes,
{\em Spectral action for Robertson-Walker metrics}, J. High Energy Phys. 
2012, no. 10, 101 [29 pages]

\bibitem{CCM}
A.~Chamseddine, A.~Connes, M.~Marcolli, {\em 
Gravity and the standard model with neutrino mixing}, 
Advances in Theoretical and Mathematical Physics, 11 (2007) 991--1090.

\bibitem{CIL}  E.~Christensen, C.~Ivan, M.L.~Lapidus, {\em
Dirac operators and spectral triples for some fractal sets built on curves}, Adv. Math. 217 (2008), 
N.1, 42--78.

\bibitem{CIS} E.~Christensen, C.~Ivan, E.~Schrohe,
{\em Spectral triples and the geometry of fractals}, J. Noncommut. Geom. 6 (2012), no. 2, 249--274.

\bibitem{Caill}  S.~Caillerie, M.~Lachi\`eze-Rey, J.P.~Luminet, R.~Lehoucq, A.~Riazuelo, J.~Weeks, 
{\em A new analysis of
the Poincar\'e dodecahedral space model}, Astron. and Astrophys. 476 (2007) N.2, 691--696.

\bibitem{ChiDo} M.~Ching, J.R.~Doyle, 
{\em Apollonian circle packings of the half-plane}, 
J. Comb. Vol.3 (2012) N.1, 1--48. 

\bibitem{CoS3} A.~Connes,  {\em Geometry from the spectral point of
view}.  Lett. Math. Phys. 34 (1995), no. 3, 203--238.

\bibitem{Co-rec} A.~Connes, 
 {\em On the spectral characterization of manifolds}, J. Noncommut. Geom. 7 (2013) N.1, 1--82. 
 
 \bibitem{Co-qgr} A.~Connes, 
 {\em Cyclic cohomology, noncommutative geometry and quantum group symmetries},
 in ``Noncommutative Geometry" (S.~Doplicher, R.~Longo, Eds.) Lecture Notes in
 Mathematics, Vol.1831, pp.~1--71, Springer, 2004.
 
 \bibitem{CoMa-book} A.~Connes, M.~Marcolli,  {\em  Noncommutative Geometry, 
 Quantum Fields, and Motives}, Colloquium Publications, Vol.55, American Mathematical Society, 2008.
 
 \bibitem{CoMa} A.~Connes, M.~Marcolli,  {\em A walk in the noncommutative garden}, in
 ``An invitation to Noncommutative Geometry" (M.~Khalkhali, M.~Marcolli, Eds.) 
 pp.~1--128, World Scientific, 2006.
 
 \bibitem{CoMo} A.~Connes, H.~Moscovici, {\em The local index formula in noncommutative geometry},
 Geometric and Functional Analysis GAFA, Vol.5 (1995) N.2, 174--243.
 
 \bibitem{Dunne} G.V.~Dunne, {\em Heat kernels and zeta functions on fractals}, Journal of Physics A:
 Mathematical and Theoretical, Vol.45 (2012) N.37, 374016 [22 pages]
 
 \bibitem{EIS} M.~Eckstein, B.~Iochum, A.~Sitarz, {\em Heat trace and spectral action on the 
 standard Podle\'s sphere}, Communications in Mathematical Physics, Vol.332 (2014)
 N.2, 627--668.
 
 \bibitem{EckZa} M.~Eckstein, A.~Zaj\c{a}c, {\em Asymptotic and exact expansion of heat
 traces}, arXiv:1412.5100 [math-ph]
 
 \bibitem{FFM} W.~Fan, F.~Fathizadeh, M.~Marcolli, {\em 
 Spectral Action for Bianchi Type-IX Cosmological Models}, arXiv:1506.06779, to
 appear in JHEP.
 
 \bibitem{Farr} R.S.~Farr, E.~Griffiths, {\em Estimate for the fractal dimension of the
 Apollonian gasket in $d$ dimensions}, Physical Review E 81 (2010) 061403  [4 pages]      
 
 \bibitem{FGK} F.~Fathizadeh, A.~Ghorbanpour, M.~Khalkhali, 
{\em Rationality of spectral action for Robertson-Walker metrics}, 
J. High Energy Phys. 2014, no. 12, 064 [21 pages] 

\bibitem{Sylos2} A.~Gabrielli, F.~Sylos Labini, M.~Joyce, L.~Pietronero, 
{\em Statistical Physics for Cosmic Structures}, Springer, 2005.

\bibitem{LagNT} R.L.~Graham, J.C.~Lagarias, C.L.~Mallows,
A.R.~Wilks, C.H.Yan, {\em Apollonian circle packings: number theory},
J. Number Theory 100 (2003) 1--45. 

\bibitem{GLMWY} R.L.~Graham, J.C.~Lagarias, C.L.~Mallows,
A.R.~Wilks, C.H.Yan, {\em Apollonian Circle Packings: Geometry and Group Theory III.
Higher Dimensions}, Discrete Comput. Geom. 35 (2006) 37--72.

\bibitem{Hardy} G.H.~Hardy, {\em Divergent Series}, American Mathematical
Society, second edition, 1991.

\bibitem{Hawk} J.~Hawkes, 
{\em Epsilon entropy and the packing of balls in Euclidean space}, 
Mathematika 43 (1996) no. 1, 23--31. 

\bibitem{IoLe} B.~Iochum, C.~Levy, {\em Tadpoles and commutative spectral triples},
Journal of Noncommutative Geometry, Vol.5 (2011) N.3, 299--329.

\bibitem{Kigami} J.~Kigami, {\em Analysis on fractals}, Cambridge Tracts in Mathematics, 143. Cambridge University Press, 2001.

\bibitem{KoOh} A.~Kontorovich, H.~Oh, {\em Apollonian circle packings and closed horospheres on
hyperbolic 3-manifolds}, Journal of AMS, Vol 24 (2011) 603--648.

\bibitem{LMW} J.C.~Lagarias, C.L.~Mallows,
A.R.~Wilks, {\em Beyond the Descartes circle theorem}, Amer. Math. Monthly 109 (2002)
338--361.

\bibitem{Lapidus} M.L.~Lapidus, M.~van Frankenhuijsen, {\em Fractal geometry, 
complex dimensions and zeta functions. Geometry and spectra of fractal strings}, Second edition. Springer Monographs in Mathematics. Springer, 2013. 

\bibitem{Larman} D.G.~Larman, 
{\em On the exponent of convergence of a packing of spheres}, 
Mathematika 13 (1966) 57--59. 

\bibitem{Luminet}  J.P.~Luminet, J.~Weeks, A.~Riazuelo, R.~Lehoucq, 
{\em Dodecahedral space topology as an explanation
for weak wide-angle temperature correlations in the cosmic microwave background}, 
Nature 425 (2003) 593--595.

\bibitem{Mallo} 
C.~Mallows, {\em Growing Apollonian packings},
J. Integer Seq. 12 (2009) N.2, Article 09.2.1 [8 pages] 

\bibitem{MaPieTeh}
M.~Marcolli, E.~Pierpaoli, K.~Teh, {\em The spectral action and cosmic topology}, 
Comm. Math. Phys. 304 (2011), no. 1, 125--174

\bibitem{MaPieTeh2}
M.~Marcolli, E.~Pierpaoli, K.~Teh, {\em The coupling of topology and inflation in 
noncommutative cosmology}, Comm. Math. Phys. 309 (2012), no. 2, 341--369.

\bibitem{Moraal}
H.~Moraal, {\em Apollonian arrangements of spheres in $d$-dimensional space}, 
J. Phys. A 27 (1994) N.23, 7785--7791. 

\bibitem{MuDy} 
J.R.~Mureika, C.C.~Dyer, {\em Multifractal analysis of Packed Swiss Cheese Cosmologies},
General Relativity and Gravitation, Vol.36 (2004) N.1, 151--184.

\bibitem{OlSi} P.~Olczykowski, A.~Sitarz, {\em On spectral action over Bieberbach manifolds},
  Acta Phys. Polon. B 42 (2011) N.6, 1189--1198.

\bibitem{Peeb} P.J.E.~Peebles, {\em Principles of Physical Cosmology},
Princeton University Press, 1993.

\bibitem{Rees} M.J.~Rees, D.W.~Sciama, {\em Large-scale density inhomogeneities in
the universe}, Nature, Vol.217 (1968) 511--516.

\bibitem{Rennie} A.~Rennie, {\em Smoothness and locality for nonunital spectral triples},
K-theory, 28 (2003) N.2, 127--165.

\bibitem{Rib1} M.B.~Ribeiro, {\em On Modeling a relativistic hierarchical (fractal) 
cosmology by Tolman's spacetime. I. Theory}, The Astrophysical Journal, 388 (1992)
1--8. 

\bibitem{Rib2} M.B.~Ribeiro, {\em On modeling a relativistic hierarchical (fractal) cosmology
by Tolman's spacetime. II. Analysis of the Einstein-De Sitter model}, The Astrophysical Journal,
395 (1992) 29--33.

\bibitem{Soder} B.~S\"oderberg, {\em Apollonian tiling, the Lorentz group, and regular trees},
Phys. Rev. A, Vol.46 (1992) N.4, 1859--1866.

\bibitem{Stri} R.S.~Strichartz, 
{\em Differential equations on fractals}, 
Princeton University Press, 2006

\bibitem{Sylos} F.~Sylos Labini, M.~Montuori, L.~Pietroneo,
{\em Scale-invariance of galaxy clustering}, Phys. Rep. Vol.~293 (1998) N.~2-4,
61--226.

\bibitem{Teh} K.~Teh, {\em Nonperturbative spectral action 
of round coset spaces of $SU(2)$}, J. Noncommut. Geom. 7 (2013), no. 3, 677--708.


\end{thebibliography}
\end{document}